\documentclass[longauth]{aa} 
\usepackage{graphicx}
\usepackage[varg]{txfonts}
\usepackage{color}
\usepackage{ulem}
\usepackage{natbib}
\bibpunct{(}{)}{;}{a}{}{,} 

\def\kms{\,{\rm km\, s}^{-1}}

\def\kpc{\,{\rm kpc}}
\def\EW{{\rm EW}}

\begin{document}
  \title{Discovery of a rich proto-cluster at $z=2.9$ and 
  associated diffuse cold gas in the VIMOS Ultra-Deep Survey (VUDS)\thanks{Based 
  on data obtained with the European Southern Observatory Very Large
Telescope, Paranal, Chile, under Large Program 185.A-0791.
}}

\titlerunning{Proto-cluster at $z=2.9$ in VUDS}

   \author{O. Cucciati\inst{1,2} 
          \and G. Zamorani\inst{2}
	  \and B. C. Lemaux\inst{3}
          \and S. Bardelli\inst{2}
          \and A. Cimatti\inst{1,2}
	  \and O. Le F\`evre\inst{3}
	  \and P. Cassata\inst{3,4}
	  \and B. Garilli\inst{5}
	  \and V. Le Brun\inst{3}
	  \and D. Maccagni\inst{5}
	  \and L. Pentericci\inst{6}
	  \and L. A. M. Tasca\inst{3}
	  \and R. Thomas\inst{3}
	  \and E. Vanzella\inst{2}
	  \and E. Zucca\inst{2}
	  \and R. Amorin\inst{6}
	  \and P. Capak\inst{7}
	  \and L. P. Cassar\`a\inst{5}	  
          \and M. Castellano\inst{6}
	  \and J.G. Cuby\inst{3}
	  \and S. de la Torre\inst{3}
	  \and A. Durkalec\inst{3}
	  \and A. Fontana\inst{6}
	  \and M. Giavalisco\inst{8}
	  \and A. Grazian\inst{6}
	  \and N. P. Hathi\inst{3}
	  \and O. Ilbert\inst{3}
	  \and C. Moreau\inst{3}
	  \and S. Paltani\inst{9}
	  \and B. Ribeiro\inst{3}
	  \and M. Salvato\inst{10}
	  \and D. Schaerer\inst{11,12}
	  \and M. Scodeggio\inst{5}
	  \and V. Sommariva\inst{1,6}
	  \and M. Talia\inst{1}
	  \and Y. Taniguchi\inst{13}
	  \and L. Tresse\inst{3}
	  \and D. Vergani\inst{14,2}
	  \and P.W. Wang\inst{3}
	  \and S. Charlot\inst{15}
	  \and T. Contini\inst{12}
	  \and S. Fotopoulou\inst{9}
	  \and C. L\'opez-Sanjuan\inst{16}
	  \and Y. Mellier\inst{15}
	  \and N. Scoville\inst{7}}
   \offprints{Olga Cucciati (olga.cucciati@oabo.inaf.it)}
\institute{
University of Bologna, Department of Physics and Astronomy (DIFA), V.le Berti Pichat, 6/2 - 40127, Bologna
\and 
INAF--Osservatorio Astronomico di Bologna, via Ranzani,1, I-40127, Bologna, Italy
\and 
Aix Marseille Universit\'e, CNRS, LAM (Laboratoire d'Astrophysique de Marseille) UMR 7326, 13388, Marseille, France
\and 
Instituto de Fisica y Astronomia, Facultad de Ciencias, Universidad de Valparaiso, Gran Bretana 1111, Valparaiso, Chile
\and 
INAF--IASF, via Bassini 15, I-20133,  Milano, Italy
\and 
INAF--Osservatorio Astronomico di Roma, via di Frascati 33, I-00040,  Monte Porzio Catone, Italy
\and 
Department of Astronomy, California Institute of Technology, 1200 E. California Blvd., MC 249-17, Pasadena, CA 91125, USA 
\and 
Astronomy Department, University of Massachusetts, Amherst, MA 01003, USA 
\and 
Department of Astronomy, University of Geneva, ch. d'\'Ecogia 16, CH-1290 Versoix
\and 
Max-Planck-Institut f\"ur Extraterrestrische Physik, Postfach 1312, D-85741, Garching bei M\"unchen, Germany
\and 
Geneva Observatory, University of Geneva, ch. des Maillettes 51, CH-1290 Versoix, Switzerland
\and 
Institut de Recherche en Astrophysique et Plan\'etologie - IRAP, CNRS, Universit\'e de Toulouse, UPS-OMP, 14, avenue E. Belin, F31400 Toulouse, France
\and 
Research Center for Space and Cosmic Evolution, Ehime University, Bunkyo-cho 2-5, Matsuyama 790-8577, Japan
\and 
INAF--IASF Bologna, via Gobetti 101, I--40129,  Bologna, Italy
\and 
Institut d'Astrophysique de Paris, UMR7095 CNRS, Universit\'e Pierre et Marie Curie, 98 bis Boulevard Arago, 75014 Paris, France
\and 
Centro de Estudios de F\'isica del Cosmos de Arag\'on, Teruel, Spain}
             
 
  \abstract{High-density environments are crucial places for studying the
   link between hierarchical structure formation and stellar mass
   growth in galaxies. In this work, we characterise a massive
   proto-cluster at $z=2.895$ that we found in the COSMOS field using
   the spectroscopic sample of the VIMOS Ultra-Deep Survey
   (VUDS). This is one of the rare structures at $z\sim 3$ not
   identified around an active galactic nucleus (AGN) or a radio galaxy, 
   thus it represents an
   ideal laboratory for investigating the formation of galaxies in dense
   environments. The structure comprises 12 galaxies with secure
spectroscopic redshift in an area of $\sim7'\times8'$, in a total $z$
range of $\Delta z = 0.016$. The measured galaxy number overdensity is
$\delta_g =12 \pm 2$. This
overdensity has a total mass of $M\sim8.1\times10^{14}M_{\odot}$ in a
volume of $13\times15\times17$ Mpc$^3$. Simulations 
indicate that such an overdensity at $z\sim2.9$ is a proto-cluster, 
which will collapse in a cluster of total mass 
$M_{z=0}\sim2.5 \times 10^{15} M_{\odot}$ at $z=0$, i.e. a massive 
cluster in the local Universe.  We analysed 
the properties
of the galaxies within the overdensity, and we compared them with a
control sample at the same redshift but outside the
overdensity. We could not find any
statistically significant difference between the properties (stellar
mass, star formation rate, specific star formation rate, NUV-r and r-K colours) 
of the galaxies inside and outside the
overdensity, but this
result might be due to the lack of statistics or possibly to the 
specific galaxy population sampled by VUDS, which could be less  
affected by environment than the other populations not 
probed by the survey. The stacked spectrum of galaxies in the background of the
overdensity shows a significant absorption feature at the 
wavelength of Ly$\alpha$ redshifted at $z=2.895$
($\lambda=4736$\AA), with a rest frame equivalent width ($EW$) of $4\pm 1.4$\AA. Stacking 
only background 
galaxies without
intervening sources at $z\sim2.9$ along their line of sight, we find
that this absorption feature has a rest frame EW of $10.8\pm3.7$\AA,
with a detection S/N of $\sim4$. We verify that this measurement is
not likely to be due to noise fluctuations. These EW values imply a 
high column density (N(HI)$\sim3-20\times10^{19}$cm$^{-2}$), 
consistent with a scenario where
such absorption is due to intervening cold streams of gas  that are
falling into the halo potential wells of the proto-cluster galaxies.
Nevertheless, we cannot rule out the hypothesis that this absorption line 
is related to the diffuse gas within the
overdensity.}


   \maketitle

\section{Introduction}\label{intro}

The detection and study of (proto) clusters at high redshift is 
important input for cosmological models, and these high-density
environments are crucial places for studying the link between
hierarchical structure formation and stellar mass growth in galaxies
at early times. The earlier the epoch when an overdensity is
detected, the more powerful the constraints on models of galaxy
formation and evolution because of the shorter cosmic time over which
physical processes have been able to work. Specifically, in high-redshift 
($z\gtrsim1.5-2$) (proto) clusters the study of how environment affects 
the formation  and evolution of galaxies is particularly effective, 
because galaxies had their peak of star formation  at such 
redshifts \citep{madau1996,cucciati12sfrd}.

However, the sample of high-redshift ($z\gtrsim 1.5$) structures
detected so far is still limited, and it is very heterogeneous,
spanning from relaxed to unrelaxed systems. Many detection techniques
have been used that are based on different (and sometimes apparently
contradicting) assumptions. For instance, the evolution of galaxies in
clusters appears to be accelerated relative to low-density
regions (e.g. \citealp{steidel05}).  As a result, while the average
SFR of a galaxy decreases with increasing local galaxy density in the
low-redshift Universe, this trend should reverse at earlier times,
with the SFR increasing with increasing galaxy density
\citep{cucciati2006,elbaz2007}. Indeed, some (proto-) clusters have
been identified at high redshift as overdensities of star-forming
galaxies \citep{capak11}, such as Ly$\alpha$ emitters
\citep{steidel00,Ouchi03,Ouchi05,lemaux09} and H$\alpha$ emitters
\citep{hatch11b}. At the same time, in some high-$z$ overdensities
an excess of massive red galaxies has also been observed
(e.g. \citealp{kodama07,spitler12}), and other dense structures have
been identified via a red-sequence method (e.g. \citealp{andreon09})
or via an excess of IR luminous galaxies
(e.g. \citealp{gobat11,stanford12}) or LBGs
(e.g. \citealp{toshikawa12}). 

High-$z$ overdensities have also been identified by using other
observational signatures, for instance with the 
Sunyaev-Zeldovich effect \citep{sunyaev_zeldovich1972,sunyaev_zeldovich1980}
as in \cite{foley11_SZ}, or searching for diffuse X-ray emission
(e.g. \citealp{fassbender11}), or looking for photometric redshift
overdensities in deep multi-band surveys
\citep{castellano07,salimbeni09,scoville13,chiang14_cosmos}. Moreover,
assuming a synchronised growth of galaxies with that of their
super-massive black holes, high-redshift proto-structures have been
searched for around AGNs (e.g. \citealp{miley04}) and radio galaxies
(e.g. \citealp{pentericci00,matsuda09,galametz12}), even if an excess
of galaxies around high-$z$ QSOs has not always been found (see
e.g. \citealp{decarli12}). However, this approach could introduce
unknown selection effects, for example those due to the influence of
powerful radio galaxies on the surrounding environment.
The study of proto-structures selected only on the basis of the
redshift distribution of its members is more likely to offer an
unbiased view of high-density environments at high redshift and allow
a comparison with the habitat of radio galaxies and
quasars. Nevertheless, it is necessary to obtain spectroscopic
redshifts of member galaxies, which is a costly observational task at
high redshifts. Spectroscopic surveys conducted with visible
wavelength spectrographs will observe the UV rest frame light of
galaxies at redshifts $z>2$, and therefore be mostly sensitive to
star-forming galaxies.

Although the sample of (proto) clusters at $z>1.5$ is increasing in
number, it is a heterogeneous data set. This inhomogeneity prevents
using it to assess the abundance of clusters at such redshifts,
which could be used to constrain cosmological
parameters (e.g. \citealp{borgani1999,ettori09}). \cite{chiang13_sim} 
have recently made an attempt to find a
common parameter to group and analyse the known overdensities at high
$z$. They used simulations to study the probability of given
overdensities at $z=2-5$ to collapse in bound clusters at $z=0$, and,
in case of collapse, the mass at $z=0$ ($M_{z=0}$) of such
clusters. They also give prescriptions for computing $M_{z=0}$ using the
overdensity of the proto-cluster within a given volume. Following
their own prescriptions, in a second work \citep{chiang14_cosmos} they
perform a homogeneous search for overdensities using the photometric
redshifts in the COSMOS field. We come back to their analysis in
the following sections.

The discovery and study of an overdensity at high $z$ also naturally
addresses how a dense environment affects galaxy
formation and evolution. Galaxies can build their stellar masses via 
abrupt processes like mergers, which in some cases produce an 
increase in mass up to a factor of two or so, or via more continuous 
processes based on in-situ star formation. At the same time, other 
physical processes are likely at work to quench star formation (such as  
AGN and SNe feedback), and some of these processes are particularly 
effective in high-density environments, where the gas reservoirs in 
galaxies can be stripped during interactions with the intra-cluster 
medium (ICM). 

The relative role of all these processes as a function of cosmic time
is still a matter of debate. In recent years, many observational
studies have focused on analysing the merger rate. If at $z<1$ the
evolution of merger rate is quite well constrained for both major and
minor mergers (i.e. with a luminosity/mass ratio greater or less than
$\sim1/4$, see e.g. \citealp{deravel09} and
\citealp{lopez_sanjuan11_vvds}), at $z>1$ observational results 
still show a large scatter (see e.g. \citealp{lopez_sanjuan13_massiv}
and \citealp{tasca14_merger} for the most recent studies).  On the
side of stellar mass growth via smooth star formation, some
theoretical models support a scenario where massive ($M_{baryon} \sim
10^{11} M_{\odot}$ ) galaxies at $z \sim 2 - 3$ are efficiently fed by
narrow, cold (e.g. $T\sim 10^4$ K), intense, partly clumpy, gaseous
streams that penetrate the shock-heated halo gas into the
inner galaxy with rates of the order of 100 $M_{\odot} {\rm yr}^{-1}$
. These streams can grow a dense, unstable, turbulent disc with a
bulge and trigger rapid star formation
(e.g. \citealp{keres05,dekel09}). Observational evidence of gas
accretion is still limited \citep{giavalisco11,bouche13}, and further
studies are needed to support this scenario. Simulations
\citep{kimm11} show that the covering fraction of dense cold gas is
larger in more massive haloes, suggesting that the best environment for 
testing the cold flow accretion scenario are high-redshift over-dense
regions.

In this paper, we present the discovery of an overdensity at
$z\sim2.9$ in the COSMOS field, detected in the deep spectroscopic
survey VUDS (VIMOS Ultra-Deep Survey). In Sect.~\ref{data} we describe our data. In
Sect.~\ref{overdensity} we describe the overdensity and compute the
total mass that it comprises, and also its possible evolution to
$z=0$. Section \ref{background} shows the search for diffuse cold gas in
the overdensity, as inferred by absorption lines in the spectra of
background galaxies. In Sect.~\ref{galaxies} we analyse the 
properties of the galaxies in the overdensity and contrast them to a
sample of galaxies outside the structure at a similar
redshift. Finally, in Sect.~\ref{discussion} we discuss our results
and summarise them in Sect.\ref{summary}.

We adopt a flat $\Lambda$CDM cosmology with
$\Omega_m=0.27$, $\Omega_{\Lambda}=0.73$, and $H_0=70$
km~s$^{-1}$~Mpc$^{-1}$. Magnitudes are expressed in the AB system.


\section{Data}\label{data}

The VUDS survey is fully described in \cite{lefevre14}, so we give 
only a brief summary here. VUDS is a spectroscopic survey using VIMOS on
the ESO-VLT \citep{lefevre2003}, targeting mainly $z > 2$ galaxies in one
square degree in three fields: COSMOS, ECDFS, and VVDS-2h. Spectroscopic
targets have been mainly selected based on a photometric redshift 
($z_p$) cut. Photometric redshifts are derived with the code {\it Le
Phare\footnote{http://www.cfht.hawaii.edu/$\sim$arnouts/LEPHARE/lephare.html}}
\citep{arnouts99,ilbert2006_pz} using the multi-wavelength photometry
available in the survey fields, and they have an accuracy of $\sigma_{zp}
\simeq 0.04(1+z)$ for magnitudes $i_{AB} \leq 25$ in the COSMOS field
(see \citealp{ilbert2013}). The primary criterion for target selection 
in VUDS is for targets to satisfy $z_p + 1\sigma_{zp}> 2.4$ and 
$i_{AB} \leq 25$. To account 
for degeneracies in the $z_p$ computation, in this criterion $z_p$  
could be   either the first or second peak of the $z_p$ probability 
distribution function.

The VIMOS spectra have been observed with 14h integrations with the
LRBLUE (R=230) and LRRED (R=230) grisms, covering a combined
wavelength range 3600 $< \lambda< 9350\AA$. This integration time 
allows reaching S/N $\sim 5$ on the continuum (at $\lambda\sim8500$\AA) 
for galaxies with $i_{AB} = 25$, and S/N $\sim 5$ for an emission line with a
flux $F = 1.5\times10^{-18}erg/s/cm^{2}/\AA$.

The spectroscopic redshift accuracy with
this setup is $\sigma_{zs}= 0.0005 (1 + z)$ \citep{lefevre2013a},
corresponding to $\sim150\kms$.  Data reduction, redshift measurement, 
and assessment of the reliability of measured redshift are described
in full detail in \cite{lefevre14}. In brief, data are reduced 
within the VIPGI environment \citep{scodeggio05}, and then the 
spectroscopic redshifts are measured with the software EZ \citep{garilli10}. EZ 
is based on the cross-correlation with templates. We used templates 
derived from previous VIMOS observations of the VVDS and zCOSMOS surveys. A 
first redshift measurement is obtained by a blind EZ run, then 
two different team members inspect the result, separately, and 
modify it if needed. Finally, the two measurements are compared 
and the two measurers provide, in agreement, a single final measurement.

During the measurement process, a reliability flag is also assigned to
each measured redshift by the two measurers, namely flag=1, 2, 3, 4, 9.
Based on previous VIMOS surveys similar to VUDS (see e.g. the VVDS
survey, \citealp{lefevre2013a}), the redshifts with
flag=1, 2, 3, 4, 9 should have a probability of being right of
$\sim50,80,98,100,80$\%, respectively. The precise assessment of such
probability values for VUDS is ongoing. There are also objects with flag=0,
i.e. when no redshift could be assigned. Brighter than $i_{AB}=25$,
the fraction of VUDS targets with a reliable redshift measurement
(i.e. flag=2,3,4,9) is $\sim74$\%.

\subsection{Ancillary photometric data and rest-frame galaxy properties}\label{data_SED}

In addition to the VIMOS spectroscopic data, a large set of imaging
data is available in the three fields. In particular, the COSMOS field
\citep{scoville2007_COSMOS} has a full coverage with the HST-ACS F814W
filter \citep{koekemoer07_HST} and includes, among other data, $BVriz$
photometry from Subaru \citep{capak2007_photcat,taniguchi07}, and the more
recent $YJHK$ photometry from the UltraVista survey
\citep{mccracken12}.

Absolute magnitudes, stellar masses ($\mathcal{M}$), and star formation rates 
(SFRs) for the spectroscopic sample and for the photometric parent catalogue  
were computed using the
code {\it Le Phare} as described in \cite{ilbert2013}, using the
measured spectroscopic redshift when available.  The method is based
on a spectral energy distribution (SED) fitting technique.  
We used a template set that
comprises \cite{BC03} templates. We assumed the \cite{calzetti2000}
extinction law and included emission line contributions as described
in \cite{ilbert09}. We used a library based on a composite of delayed star 
formation histories (SFH) and exponentially declining SFHs, with nine possible $\tau$ 
values ranging from 0.1 Gyr to 30 Gyr. 

The typical statistical error on $\mathcal{M}$ and SFR is
$\sim0.15$ and $0.18$ dex, respectively, at $z\sim3$, i.e. the
redshift of interest in this work. Moreover, we tested which is the
impact of the choice of different SFHs on $\mathcal{M}$ and SFR. We
recomputed the SED fitting another two times, in one case using only
delayed SFHs and then using only exponentially declining SFHs. We find
that the difference on the derived $\mathcal{M}$ among the two SFH
sets is negligible, as also found in \cite{ilbert2013}. In contrast,
there is a systematic offset of $\sim0.05$ dex between the SFR based
on delayed SFHs and  the SFR based
on exponentially declining SFHs, with a scatter
around the offset of about 0.03 dex.

The absolute
magnitude computation is optimised using the full information given
by the multi-band photometric data described above. To limit the template 
dependency, absolute magnitudes in each band are based 
on the observed magnitude in the band that, redshifted in the observer 
frame, is the closest to the given absolute magnitude band  
\citep[see Appendix A.1 in][]{ilbert2005}.  Because of this adopted method, 
the biggest source of uncertainty on the absolute magnitude computation 
is the error on the closest observed magnitude, and there is essentially no difference 
when using different SFHs. Namely, the typical errors on the 
$NUV$, $r$, and $K$ absolute magnitudes used in this paper 
(see Sect.\ref{galaxies}) are of 
the order of 0.06, 0.1, and 0.15 magnitudes, respectively.

Given the main selection of our sample, i.e. $i_{AB} \leq 25$, VUDS
mainly probes (relatively) star-forming galaxies at the redshift of
interest for this paper ($z\sim3$), where the $i$-band corresponds to
the rest frame ultra-violet emission. For instance, we verified that
our sample, at $z\sim3$, does not probe in a complete way the range of
specific SFR (sSFR) at $\log_{10}(sSFR[yr^{-1}]) \lesssim -9$ with
respect to a catalogue selected in $K$-band (namely, $K\leq 24$, like
the one in \citealp{ilbert2013}). Galaxies with such
a low sSFR, which are missed by our survey, are generally the most
massive ones ($\log_{10}(M/M_{\odot})\gtrsim 10.5$).

\begin{figure} \centering
\includegraphics[width=8cm]{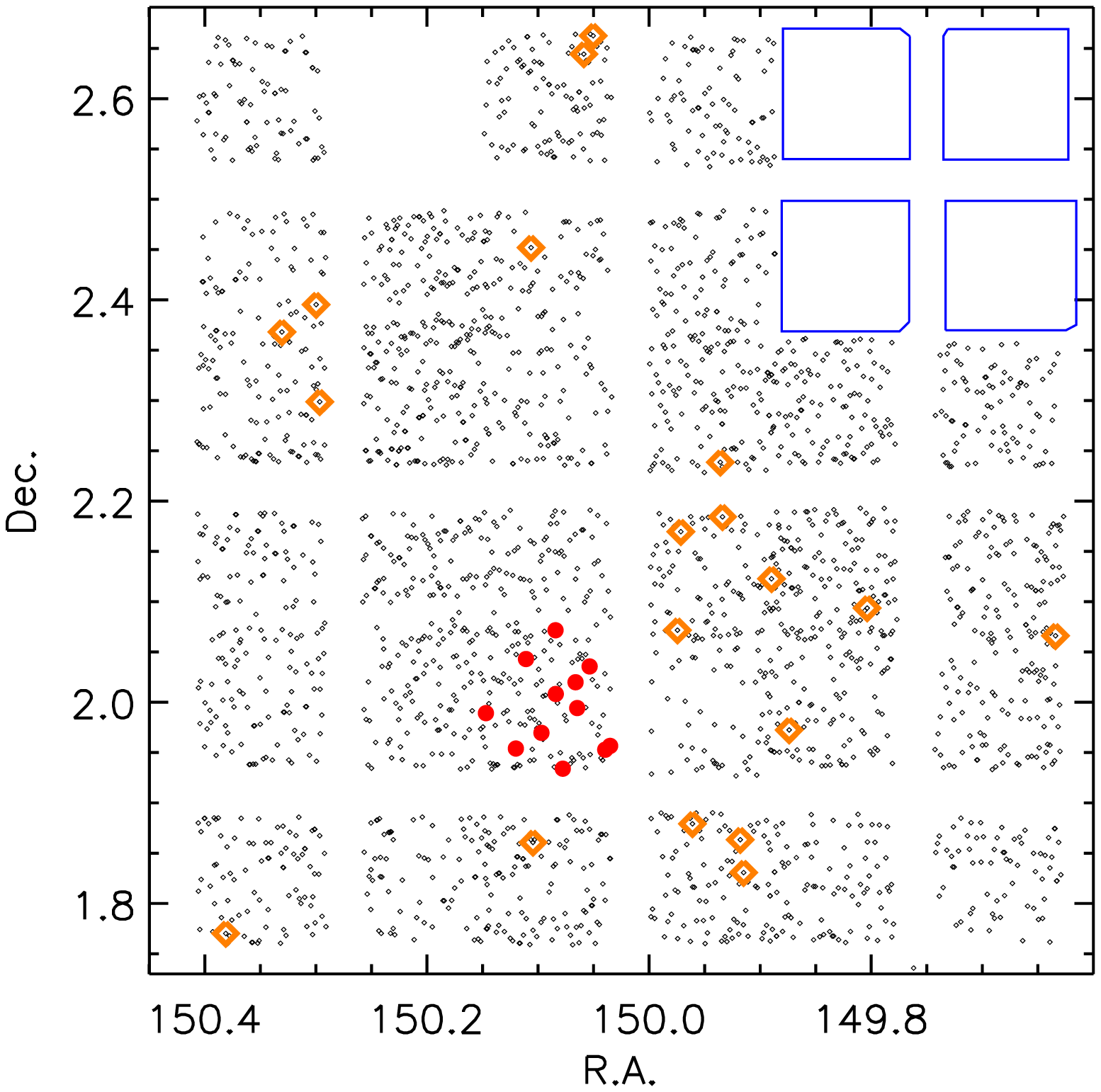}
\includegraphics[width=8cm]{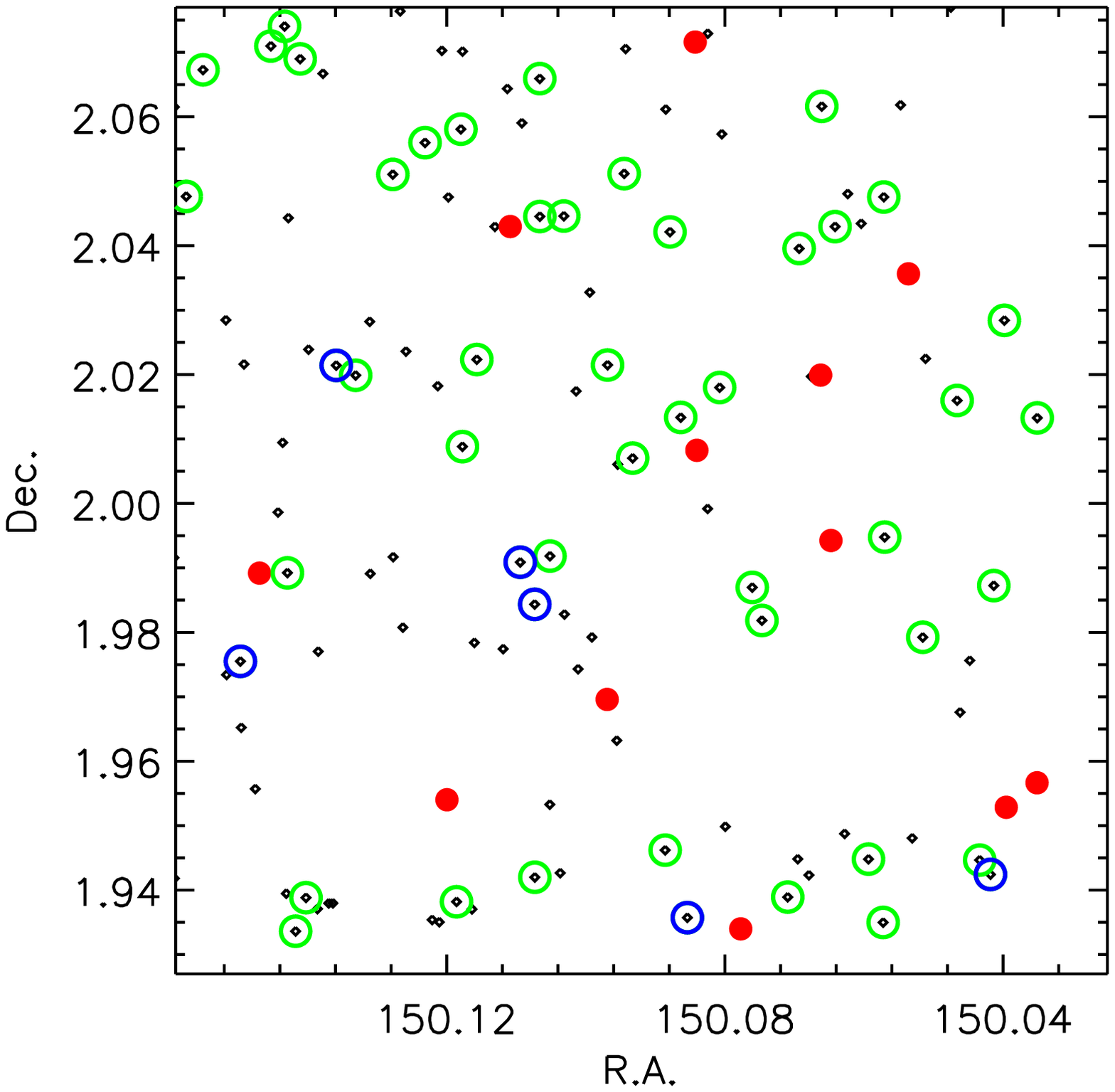} 
\caption{{\it Top:} right ascension (RA) and declination (Dec) distribution of VUDS galaxies in the
COSMOS field (dark dots). Red filled circles: VUDS galaxies within
$2.8858\leq z \leq 2.9018$ in the VIMOS quadrant where the structure
has been found (P2Q1). Orange open diamonds: other VUDS galaxies
within $2.8858\leq z \leq 2.9018$ but outside P2Q1.  The blue
rectangles show the VIMOS footprint. {\it Bottom:} like the top panel,
but zoomed-in in the region of P2Q1. Green open circles are
VUDS galaxies at $z\geq 3$, and blue open circles are
`free-line-of-sight' VUDS galaxies at $z\geq 3$ (see text for
details).}
\label{radec}
\end{figure}

\begin{figure} \centering
\includegraphics[width=8cm]{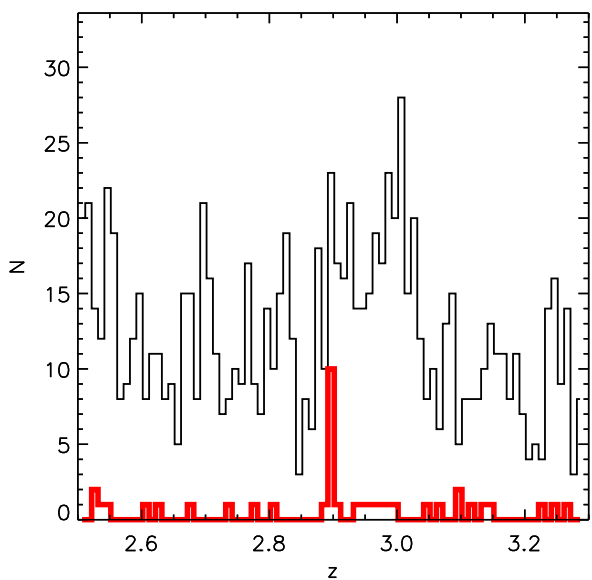}
\caption{VUDS redshift distribution in the COSMOS field (black) and in the
quadrant of the structure (red), in redshift bins of
$\Delta z=0.01$. }
\label{nz} 
\end{figure}


\section{The spectroscopic overdensity}\label{overdensity}

Using the VUDS spectroscopic data, we discovered a galaxy overdensity
in the COSMOS field at $z = 2.895$. It comprises 12 galaxies with
reliable spectroscopic redshifts (1 galaxy with flag=2, 11 galaxies
with flag 3 or 4) in a very narrow redshift bin ($\Delta z = 0.016$,
namely $2.8858\leq z \leq 2.9018$) in a region covered by a single
VIMOS quadrant. The distribution in right ascension (RA) and declination 
(Dec) of the VUDS galaxies in
the COSMOS field and, in particular, of those at $2.8858\leq z \leq
2.9018$ is shown in the top panel of Fig.~\ref{radec}. The panel also shows 
the VIMOS typical footprint, made by four
quadrants. Namely, 11 out of 12 galaxies have been detected in
quadrant Q1 of the pointing F51P002 (P2Q1 from now on), and 1 in the
adjacent quadrant F51P005 Q4 (same $RA$ as P2Q1 but at higher
$Dec$). The two quadrants overlap, so this twelfth galaxy is
within the area covered by P2Q1. From now on, we  consider this
galaxy as part of P2Q1. One VIMOS quadrant covers a region of
$\sim7'\times8'$, which corresponds, at $z\sim2.9$, to $\sim13\times15$
Mpc$^2$ comoving, or $\sim3.4\times3.9$ Mpc$^2$ physical.

Figure \ref{nz} shows the VUDS redshift distribution in the COSMOS field
and the one in P2Q1, both made using only reliable redshifts (flag=2,
3, 4, and 9), in redshift bins of $\Delta z=0.01$. There is an evident
peak of ten galaxies in P2Q1 at $z=2.895$, with two other galaxies in
the two adjacent redshift bins, for a total of 12 galaxies in $\Delta
z=0.016$. In all the quadrants in the three VUDS fields, 
in the range $2.5<z<3.3$, the median number of spectroscopic galaxies 
with reliable redshifts falling in such narrow $z$ bins is 0.   
In only 1.5\% of the cases these narrow bins comprise more 
than three galaxies, and always less than seven,  with the exception of the 
structure described in this paper. The few cases with 5-6 spectroscopic 
galaxies within $\Delta
z=0.016$ will be inspected in detail 
in a future work.  Using the 12 galaxies in the overdensity, we measured a
velocity dispersion along the line of sight of $\sigma_{los}=299 \pm
70 \kms$, but we cannot assume that the galaxy velocities in our
overdensity follow a Gaussian distribution. A more appropriate
estimator of the dispersion of a non-Gaussian (or contaminated
Gaussian) distribution is the ``biweight method'', which has been
proven to also be very robust in case of small number statistics (see
{\it case C} in \citealp{beers1990}, and references therein). With
this method we estimate $\sigma_{los}=270 \pm 80 \kms$. We retain
this second estimation as the velocity dispersion of the galaxies in
the overdensity.

We notice that we did not find any broad line AGN among the 12 VUDS
galaxies. In the Chandra-COSMOS \citep{elvis09_chandra} and XMM-COSMOS
\citep{cappelluti07_XMM} catalogues, it has not been detected any
extended nor point-like source that matches the galaxies in the
structure (Cappelluti, private communication), and we found only one
match with the sources in the Herschel PEP \citep{lutz11_PEP}
catalogue.

We also explored whether the overdensity discovered in the VUDS
spectroscopy survey is detectable using photometric redshifts, i.e. in
the $RA$-$Dec$-$z_p$ space.  To do this, we applied both the Voronoi
Tessellation algorithm \citep{voronoi1908} and the DEDICA algorithm
\citep{bardelli98}. The two methods give equivalent results, so 
we discuss only the DEDICA results.  DEDICA is an algorithm
based on an adaptive kernel method estimate of the density field, also
estimating the significance of the detected structures. 
We used the VUDS photometric parent catalogues, with the most recent 
photometric redshifts obtained by also using the YJHK bands of the 
UltraVista survey.  To
maximise the signal, we limited the analysis of the 
photometric redshift catalogue to the $i$-band range $[24-25]$ (the
same range that comprises the 12 spectroscopic galaxies in the
overdensity) and to the redshift range $[2.72-3.19]$.  The choice of
this redshift range is due to the mean difference between the
photometric ($z_p$) and spectroscopic redshift ($z_s$) of the 12
galaxies in the overdensity. We found $\langle z_p - z_s \rangle =
0.06 \pm 0.12$. To choose the redshift range for the photometric
redshifts, first we applied this shift of 0.06 to the redshift of the
structure, then we considered a range of $\pm 1\sigma = \pm0.12$ around
this redshift. After running the algorithm on this photometric data
set, we found no significant overdensities (at $90 \%$) in the region
of our structure. We obtained the same result by increasing the photometric
redshift range up to $\pm 2\sigma$ and also without applying the
shift.

We verified with a  simple model whether our spectroscopic 
overdensity, given its characteristics, would be detectable in the 
parent photometric catalogue using photometric redshifts as 
described above.  We assumed a sampling rate 
of $\sim25$\% in P2Q1 for the spectroscopic galaxies in the 
$i$-band range $[24-25]$ (see the following section). This would 
give us $\sim50$ galaxy members in the overdensity in the full 
parent photometric catalogue. We modelled the $z_s$ of these 50 galaxies 
to be distributed with $\sigma_{los}=270 \pm 80 \kms$, centred on 
$z=2.895$. To this $z_s$ distribution, we added  a random photometric 
error extracted from a Gaussian distribution with $\sigma=0.12$ 
(found above). We repeated  this computation 1000 times, and 
each time we counted the number of galaxies within the redshift 
range  $[2.72-3.19]$, and computed the overdensity $\delta_g$  
with respect to all the galaxies in the photometric 
catalogue 
in the same redshift range (and with $i$-band magnitude in the 
range $[24-25]$). We found $\delta_g=0.62 \pm 0.1$, i.e. $\sim20$ times lower 
than the value we find using spectroscopic redshift (see next section).

We conclude that, given the characteristics of the spectroscopic 
overdensity (number of member galaxies and measured 
velocity dispersion), it is unlikely to detect it using photometric 
redshifts with a typical error like  the ones we used. In the recent works by 
\cite{scoville13} and \cite{chiang14_cosmos}, who identify overdensities 
in the COSMOS field using photometric redshifts, no overdensity at 
the position of P2Q1 at $z\sim2.9$ is detected, but they both find 
a very close overdensity at 
about the same redshift.  Namely, 
\cite{chiang14_cosmos} have published a list of proto-cluster candidates
in the COSMOS field (see their Table 1). Among these candidates, the
closest in the $RA-Dec$ plane to our proto-cluster has ID=30, with
$RA=150.009$ and $Dec=1.974$, so very close ($\sim5$') to the centre of
our structure. They assign it a photometric redshift $z_p=2.77$. Their
proto-cluster candidates are detected using a sample of galaxies 
limited in $K_s$-band ($K_s<23.4)$, with photometric redshifts with an
uncertainty of $\sigma=0.025(1+z)$. The galaxy overdensity is computed
in a redshift slice of depth $0.025(1+z)$. Considering this
uncertainty on the redshift of their proto-cluster, its redshift would
be compatible with ours ($z=2.895$) at $\gtrsim
1\sigma$.

\subsection{The galaxy density contrast}\label{gal_dens}

An approximate estimate of the overdensity of this structure can be
derived in the following way. Given the number of galaxies at
$z\sim2.9$ in the VUDS area in the COSMOS field and the ratio between
such area and the area of one quadrant, the expected number of
galaxies in a redshift bin $\Delta z = 0.016$ in one quadrant is $\sim
0.71$. Thus, the estimated galaxy overdensity in this quadrant is
$\delta_g\sim(12-0.71)/0.71 \thickapprox 16$. To compute the
uncertainty on $\delta_g$ given by the uncertainty in the counts in
the field, we perform the same computation via bootstrapping,
randomly selecting 31 quadrants (with repetitions
allowed) for 5000 times, and each time computing the estimated overdensity with respect to the
counts in the 31 selected quadrants. In this way, we obtain $\delta_g
\sim 14\pm2$. Finally, we again perform this computation, but
weighting galaxies by the spectroscopic success rate (SSR) in each
quadrant\footnote{We estimated the SSR per quadrant computing the
ratio between the number of good quality redshifts (flag 2,3,4,9) over
the total number of targeted galaxies. Quadrant by quadrant variations
are expected, due for instance to different observing conditions. },
to take into account the varying success rate in measuring the
redshifts. With this weighting scheme, we obtain $\delta_g \sim
12\pm2$. We assume that this value is our best estimate for the galaxy
overdensity. It is reasonable that, by applying the quadrant-dependent
weights, we obtain a lower overdensity than when not applying them, as P2Q1
has a spectroscopic success rate that is slightly higher than the mean.

The redshift bin $\Delta z = 0.016$ that comprises our 12 galaxies is
much smaller than the ones enclosing other overdensities at this
redshift (see e.g. the summary table in \citealp{chiang13_sim}, where
the smallest $\Delta z$ is $\sim 0.03$). We verified that our value is
not an effect of the low sampling rate of our survey, and that it
corresponds to the expected maximum extension in redshift for galaxies
at $z=2.9$ distributed with $\sigma_{los}=270 \kms$. Specifically, we
proceeded by computing a rough estimate of the total galaxy sampling
rate in P2Q1, estimated as the number of reliable redshifts over the
total number of objects in the photometric catalogue. To compute the
sampling rate we considered, in both the spectroscopic and photometric
catalogues, only galaxies with $i\leq 25$, which is the faintest
$i-$band magnitude reached by the galaxies in the overdensity, and
only galaxies with $z\geq 2.5$ ($z_s$ or $z_p$, according to the
catalogue). The result is a sampling rate of $\sim25$\%. We refer the
reader to \citealp{lefevre14} for more details on the
VUDS sampling rate. This means that we would expect $\sim
12\times4=48$ galaxies in the overdensity. From a Gaussian
distribution with $\sigma=270\kms$, we selected 48 galaxies, converted
their velocity in redshift (assuming the peak of the distribution is
at $z=2.895$), and computed the maximum range spanned in redshift
($\Delta z_{max}$). We repeated this computation 1000 times. We
averaged $\Delta z_{max}$ over the 1000 extractions and obtained a
mean value of 0.0159\footnote{We repeated this exercise varying the
sampling rate from 15\% to 35\%, and we always obtained values close
to $\Delta z_{max}=0.016$, making our result stable against the
approximated value of the sampling rate.}. This indicates that the
redshift bin of $\Delta z = 0.016$ that we use for our analysis is
consistent with the total extent in redshift of the total population
(down to $i=25$) of this overdensity, in the assumption of
$\sigma_{los}=270 \kms$.  The redshift range 2.8858-2.9018 corresponds
to $\sim17$ Mpc comoving. From now on, we set the volume (in
redshift space) containing the overdensity as
$\sim13\times15\times17=3340$ Mpc$^3$.

\subsection{The overdensity total mass}\label{mass}

In this section, we estimate the total mass contained in the volume
occupied by the overdensity. Assuming this overdensity will collapse
in a cluster (see below for details), we also estimate the total mass
that this cluster should have at $z=0$. We stress that we paid
particular attention when determining the volume in which to compute the
overdensity, especially when selecting the most appropriate
$\Delta z$ (see the previous section).  This accurate choice makes the
following computations more robust, at least for what concerns the
observed quantities to be used. 
The results of this section are discussed in Sect.~\ref{discussion}.

\subsubsection{The total mass at $z=2.9$}\label{mass_z3}

We estimate  the total mass contained in the volume
occupied by the overdensity, following \cite{steidel98}. We
used the relation

\begin{equation} \displaystyle
M=\rho_m V_{true} (1+\delta_m),
\label{eq_mass} 
\end{equation}

\noindent where $\rho_m$ is the matter density, $\delta_m$  the
matter overdensity in our proto-cluster, and $V_{true}$ the volume in
real space that encloses the proto-cluster. We computed
$V_{true}$ and $\delta_m$ as follows. First, we use the relation
between $\delta_m$ and the galaxy overdensity $\delta_{g}$:

\begin{equation} \displaystyle
1+b\delta_m = C(1+ \delta_{g}),
\label{eq_bias} 
\end{equation}

\noindent where $\delta_{g}=12$, and $b$ is the bias factor. We assume $b=2.59$,
as derived in \cite{bielby13} at $z\sim3$ for galaxies similar to
ours. The factor $C$, defined as $C=V_{app}/V_{true}$, takes the redshift 
space distortions due to peculiar velocities and
the growth of perturbations into account. The variable $V_{app}$ is the volume 
in redshift space
that encloses the proto-cluster. Assuming that the matter peak under
study is undergoing an isotropic collapse, we have the simplified
expression

\begin{equation} \displaystyle
C=1+f-f(1+\delta_m)^{1/3}. 
\label{eq_Cf} 
\end{equation}

Here we use $f(z)=\Omega_{m}(z)^{0.6}$. Solving the system of
Eqs.~\ref{eq_bias} and \ref{eq_Cf}, we find $\delta_m=2.65$ and
$C=0.60$. With these values and $V_{true}=V_{app}/C= 3340 / 0.6$
Mpc$^3$ in Eq.~\ref{eq_mass}, we obtain
$M\sim8.1\times10^{14}M_{\odot}$. A lower limit for the uncertainty on
this value is around $\sim 30$\%, computed by propagating the
Poissonian error on the galaxy counts in the structure and in the
field (used to compute the mean galaxy density). One should ideally
include at least the error on the galaxy bias, but this crude
estimate is enough for the purpose of this work. The same $\sim 30$\%
uncertainty is valid, with the same caveat, for all the mass
estimations below. We can also compute a lower limit for the volume
enclosed by the overdensity, using a depth along the line of sight
that takes the typical error in the $z_s$ measurement
into account (see Sec.~\ref{data}). The resulting volume is $\sim88$\% of the
nominal one, giving $M\sim7.1\times10^{14}M_{\odot}$.

\subsubsection{The expected mass at $z=0$}\label{mass_z0}

We estimated the current mass $M_{z=0}$ of the cluster descending
from our structure, as suggested in \cite{chiang13_sim}. As discussed
by the authors, the computation of $M_{z=0}$ allows us to have a uniform
parameter to classify the proto-clusters/structures found at higher
redshift that currently constitute a rather heterogeneous
sample. \cite{chiang13_sim} used the Millennium Run
\citep{springel2005_MILL} as the N-body dark matter simulation, with the
semi-analytical model of galaxy formation and evolution by
\cite{guo11_sam}. They studied the typical characteristics (dimension,
galaxy and matter overdensity) of proto-clusters at $z\geq2$, in
relation to the total mass that these structures would have at $z=0$
($M_{z=0}$), and they characterised their growth in size and mass with
cosmic time. To quantify the spatial extent and the size of the
structures, they defined an effective radius $R_e$ of
proto-clusters, where $R_e$ is defined as the second moment of the member
halo positions weighted by halo mass. Namely, they found that
$M_{z=0}\simeq C_e (1+\delta_{m,e}) \rho_m V_e$, where
$V_e=(2R_e)^3$ is the effective volume, and $\delta_{m,e}$ is the
total matter overdensity computed within $V_e$. The quantities 
$R_e$, $\rho$, and
$\delta_{m,e}$ are estimated at the redshift of the proto-cluster of
interest\footnote{They studied $z=2,3,4,5$.}. Here, $C_e$ is a correction 
factor of $\sim 2.5$, because they found that, irrespectively of
$M_{z=0}$ and redshift (in the range that they explored,
i.e. $2\lesssim z \lesssim 5$), $V_e$ encloses $\sim40$\% of $M_{z=0}$. They
do not need any factor to correct for redshift space distortions, since 
they work in real space. 

We computed $R_e$ using the galaxies of our
structure (instead of using haloes), weighting them by their stellar
mass (under the assumption that there is a constant ratio between the
mass of the galaxies and the mass of the hosting haloes). We find a
two-dimensional $R_e$ of $\sim 5.7$ comoving Mpc, which is a 3D $R_e$
of $\sim \sqrt{3/2}\,R_e \simeq 7$ Mpc. Our $V_e$ would then not be
too different from the volume we used above to compute $\delta_g$
($13\times15\times17$ Mpc$^3$).  Moreover, our $V_e$ is comparable to
the one used by \cite{chiang13_sim}  for their
analysis ($15^3$ Mpc$^3$).

In our computation, the main difference with \cite{chiang13_sim} is
that our $V_e$ is in redshift space, while their analysis is in real
space.  This implies that, before applying their results to our data,
we first have to transform an apparent volume ($V_{app}$) into a true
volume ($V_{true}$) via the correction factor $C$, as done above
following \cite{steidel98}.  In \cite{chiang13_sim}, $V_e$ is
$V_{true}$, so we should compute our $\delta_g$ in an apparent volume
corresponding to $V_{app}= C\,V_{true}=0.60\,V_{true}$, i.e. in a
volume that is shrunk by a factor $1.0/0.6=1.66$ with respect to
$V_e$. The choice of this smaller volume is about equivalent to
computing $\delta_g$ in the same RA-Dec area defined by $R_e$, but in
a redshift slice of $\Delta_z=0.01$ instead of $\Delta_z=0.016$. Using
$\Delta_z=0.01$, we measure $\delta_g\sim17$. We notice that in this
case, $\delta_g$ is measured within an apparent volume that in real
space would correspond exactly to $V_e$, so  we derive
$\delta_m$ using $b\delta_m = \delta_{g}$, as in
Eq.\ref{eq_bias} but without factor $C$. Using $b=2.59$ as above,
we obtain $\delta_m = 6.56$.  Using this new $\delta_m$ and
$V_{app}/C=V_e=3340$ Mpc$^3$, we obtain $M_{z=0}\sim 2.5 \times
10^{15} M_{\odot}$, with an error of at least $\sim30$\% as discussed
in Sect.~\ref{mass_z3}.

We would like to point out that, according to \cite{chiang13_sim}, a structure with
$\delta_g =12$-$17$ like ours at $z\sim2.9$ has a 100\% probability of 
evolving into a galaxy cluster at $z=0$.

\begin{figure} \centering
\includegraphics[width=9cm]{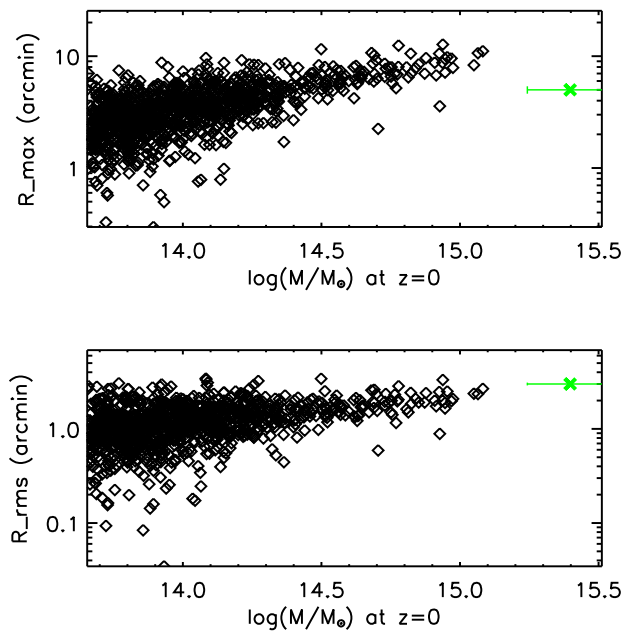} 
\caption{{\it Top.} $R_{max}$ at $z\sim2.9$ as a function of the
total mass at $z=0$ of the cluster into which the proto-cluster will
collapse. Each black diamond represents a proto-cluster at $z\sim2.9$ in the
Millennium light cones. See Sect.~\ref{mill} for the definition 
of $R_{max}$. The green cross represents our overdensity; its $R_{max}$
has to be considered a lower limit (see text for details). {\it
Bottom.} The same as in the top panel, but on the $y$-axis there is 
$R_{rms}$. See the text for the computation of $R_{rms}$.}
\label{slos_mill} 
\end{figure}

\begin{figure} \centering
\includegraphics[width=9cm]{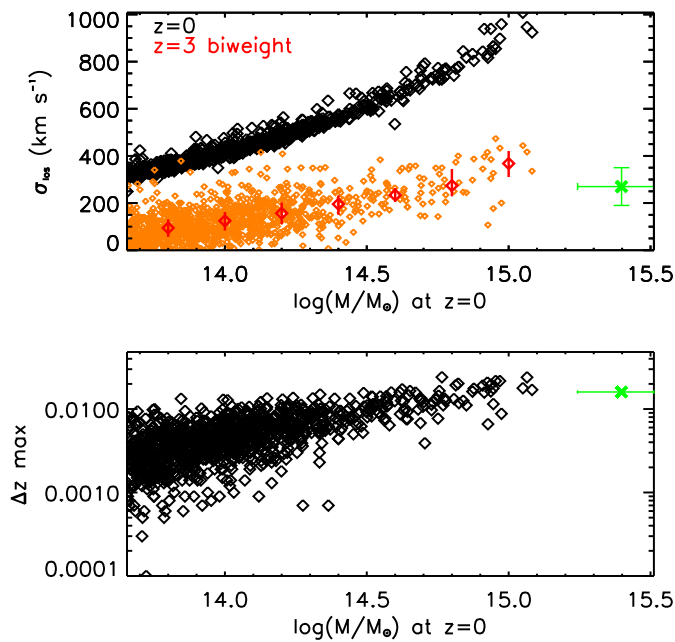} 
\caption{{\it Top.} As in
Fig.~\ref{slos_mill}, but on the $y$-axis we show the  velocity
dispersion along the line of sight ($\sigma_{los}$).  
Black points are $\sigma_{los}$
at $z=0$ as stored in the Millennium Database; orange points are
$\sigma_{los}$ computed using the redshift of the galaxies at
$z\sim2.9$ with the biweight method (see text for details); red
symbols represent the median values of the orange points, in bin of
0.2 in $\log(M/M_{\odot})$, and their bars show the 25\% and 75\% of
the distribution. The green cross represents the velocity dispersion of our
overdensity, as computed at the beginning of Sect.~\ref{overdensity}.
{\it Bottom.} The same as in the top panel, but on the $y$-axis there is
the maximum separation in redshift at $z\sim2.9$ of the galaxies in
each proto-cluster.  The green cross represents the $\Delta z_{max}$ 
of our overdensity (see Sect.~\ref{overdensity}).}
\label{zsep_mill} 
\end{figure}

\subsection{Comparison with the Millennium Simulation}\label{mill}

The contrast between the relatively small $\sigma_{los}\sim270\kms$
measured at $z\sim2.9$ and the estimated $M_{z=0}\sim 2.5\times
10^{15} M_{\odot}$ is apparently striking\footnote{If we use the
scaling relations derived from the virial theorem to compute the total
mass at $z\sim2.9$ of our overdensity, we obtain a total mass of
$M\sim2\times10^{13}M_{\odot}$ (following for instance
\citealp{finn05}), so much smaller than the total mass obtained following
\cite{steidel98} in Sect.~\ref{mass_z3}.  Of course, the use of these
scaling relations would imply the assumption that the overdensity is
virialised, which is probably not the case. }, but \cite{eke98}
showed that the velocity dispersion of a cluster increases as time
goes by, especially at $z<1$, and this consideration relaxes the
apparent inconsistency between $\sigma_{los,z=2.9}$ and
$M_{z=0}$. Nevertheless, the expected velocity dispersion at
$z\sim2.9$ for a cluster with $M_{z=0}\sim 10^{15}M_{\odot}$ is around
$400\kms$, which is greater than our findings.

\cite{eke98} used N-body hydrodynamical simulations of clusters
formation and evolution.  To better compare our results with
simulations, i.e. to study the galaxy distribution of proto-cluster members in
redshift space at $z=2.9$, 
we use galaxy mock catalogues suited to fitting at least the
basic observational characteristics of VUDS. In this section, by `proto-cluster' we
mean the set of galaxies that, according to the merger tree of their
hosting haloes, will be part of the same galaxy cluster at $z=0$. This
study has the double aim of i) verifying whether the galaxy distribution
in our overdensity is comparable to the ones found in the
simulation, and ii) verifying which is the best redshift bin to search
for (will-be) bound structures in galaxy redshift surveys.

We used ten independent light cones, derived by applying the
\cite{delucia_blaizot2007} semi-analytical model of galaxy evolution
to the dark matter halo merging trees of the Millennium Simulation
\citep{springel2005_MILL}. These light cones are limited at
$I_{AB}=25$, corresponding to the faintest magnitude of the galaxies
in our overdensity, and cover an area of $2\times2$deg$^2$ each. The
galaxy position along the line of sight includes the cosmological
redshift and the peculiar velocity.

In each of these light cones, we selected all galaxies in the same
snapshot at $z\sim2.9$. For these galaxies, it was possible to extract 
the ID of the cluster their descendants would have belonged to at
$z=0$ from the Millennium
Database\footnote{http://www.mpa-garching.mpg.de/galform/virgo/millennium/}. In this way, we grouped all galaxies at $z\sim2.9$ according to
the cluster membership at $z=0$ of their descendants. This means that,
for each cluster at $z=0$, it is possible to study the 3D distribution
and overdensity of the galaxies at $z\sim2.9$ that will collapse in
it by $z=0$.

We note that we did not apply any algorithm to identify clusters,
but we simply used the identification (ID) provided in the Millennium Database to
identify them (namely, their `friend-of-friend' halo ID). The
Millennium Database provides, for each cluster, its total mass and its
1D velocity dispersion.  We will call this mass $M_{0,sim}$ and the 1D
velocity dispersion $\sigma_{0,sim}^{1D}$.  For this
study we considered only clusters with $M_{0,sim}\gtrsim 10^{13.5} M_{\odot}$. The
highest $M_{0,sim}$ reached in the used light cones is $M_{0,sim}\sim 1.2\times
10^{15} M_{\odot}$.

For each cluster, we counted its corresponding galaxies at $z\sim2.9$
($N_{z=2.9}^{gal}$) and measured their median $RA$ and $Dec$ position
and their angular 2D distance ($R$) from this median point. For each cluster 
we retained the maximum value of $R$ ($R_{max}$) and the
rms of $R$ ($R_{rms}$).  We also computed the maximum extent in
redshift ($\Delta z_{max}$) covered by the member galaxies. We
verified that, for clusters with $M_{0,sim}\sim10^{15} M_{\odot}$, we
have $N_{z=2.9}^{gal}\sim40-50$ (down to $I_{AB}=25$). This number is in good
agreement with the number of spectroscopic galaxies in our structure (12), 
which corresponds to an expected total number of $\sim50$ if the sampling 
rate is $\sim 25$\%, as discussed in Sect.~\ref{gal_dens}.

To avoid boundary effects on the computation of $R_{max}$ and $R_{rms}$, 
we used only the clusters for which the median 
$RA$ and $Dec$ of members at $z\sim2.9$ is far at least 20 arcmin from the 
light cones boundaries.

The quantities $R_{max}$ and $R_{rms}$ are shown in
Fig.~\ref{slos_mill}, as functions of $M_{0,sim}$, together with the values for
our overdensity. The $R_{max}$ value of our
overdensity (5 arcmin) has been computed using only the galaxies in
P2Q1. Because of this, this value should be considered a lower
limit. It is not possible to measure the real $R_{max}$, since we cannot
know if the galaxies in the overall COSMOS field in the same $\Delta
z$ as the overdensity (spanning the entire $RA$-$Dec$ range of the field)
will collapse in one single cluster at $z=0$, because their density outside
P2Q1 is consistent with the field. On the contrary, thanks to our
analysis of the Millennium Simulation we expect that its $R_{max}$
will be around 10-12 arcmin (see top panel of Fig.~\ref{slos_mill}),
given its $M_{z=0}$. As a consequence, to derive the $R_{rms}$ of our
structure we used all the galaxies within a radius of 12 arcmin from
the median $RA$ and $Dec$ of the galaxies in the structure (very close
to the centre of quadrant P2Q1).

The top panel of Fig.~\ref{zsep_mill} shows the velocity dispersion
along the line of sight ($\sigma_{los}$) as a function of the total
cluster mass at $z=0$ for the clusters
in the simulation. For each cluster, we plot 
both $\sigma_{0,sim}^{1D}$ and the $\sigma_{los}$ 
computed with the biweight method using the redshift of the galaxies 
at $z\sim2.9$. When comparing black and orange points, it is evident that
$\sigma_{los}$ increases as time goes by (as already shown in
simulated clusters e.g. by \citealp{eke98}). The panel also shows that
the measured velocity dispersion of our structure is below the typical
$\sigma_{los}$ of proto-clusters in the simulation. In the plot, the
values of $\sigma_{los}$ at $z\sim2.9$ are computed with the biweight
method using all the available galaxies. For the richest
proto-clusters ($N_{z=2.9}^{gal} \geq 30$), we also measured the same
$\sigma_{los}$ but using only 12 galaxies (as in our structure). We
repeated this computation 1000 times per proto-cluster, and we
verified that the median of the distribution of such 1000
$\sigma_{los}$ is always very close to the $\sigma_{los}$ computed
using all the available galaxies, with a maximum difference of $\sim20
\kms$.

The bottom panel of Fig.~\ref{zsep_mill} is similar to the top panel in
Fig.~\ref{slos_mill}, but on the $y$-axis we plot $\Delta z_{max}$,
i.e. the maximum range in redshift covered by the galaxies at
$z\sim2.9$. This plot is particularly useful for determining which is the
more suitable redshift interval in which `proto-clusters' should be
searched for at $z\sim3$.  This figure suggests that searches for
proto-clusters at $z\sim3$ should be done in redshift bins of $\Delta
z \lesssim 0.02$ (although this result 
is based on a small sample), to which the typical redshift measurement error of
the given survey should be added in quadrature. In the case of VUDS,
the redshift measurement error is small compared to $\Delta z \sim
0.02$ ($\sigma_{zs}= 0.0005 (1 + z)$, see Sect.~\ref{data}), so its
effect is negligible.

\begin{figure*} \centering
\includegraphics[width=6cm]{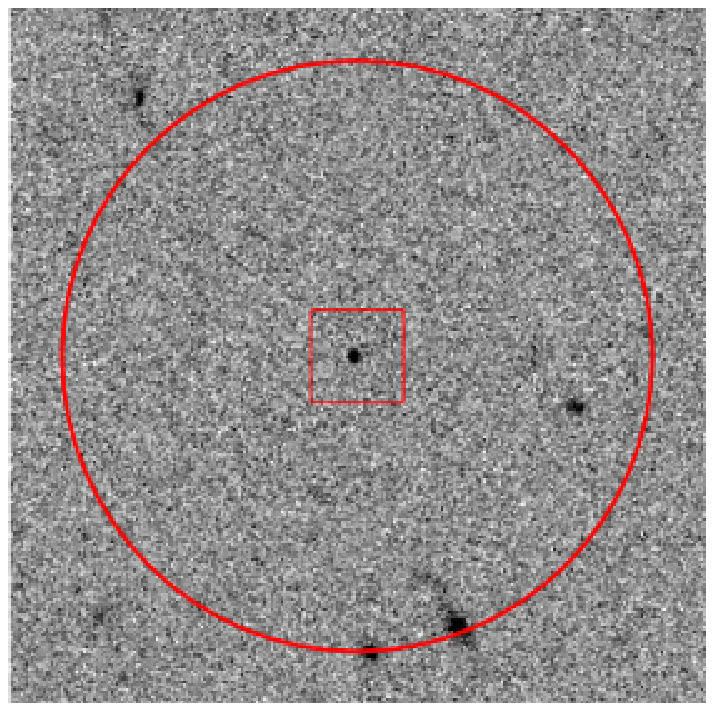}
\includegraphics[width=6cm]{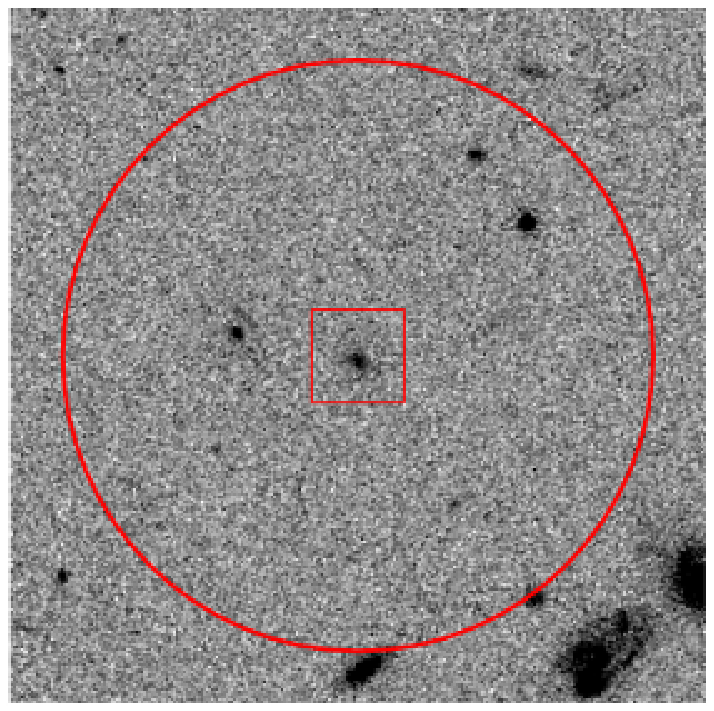}
\includegraphics[width=6cm]{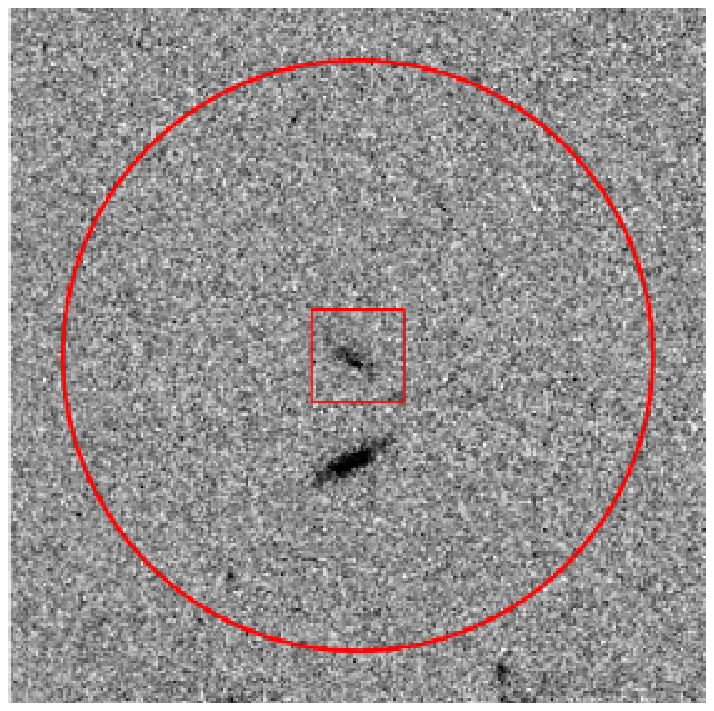}
\includegraphics[width=6cm]{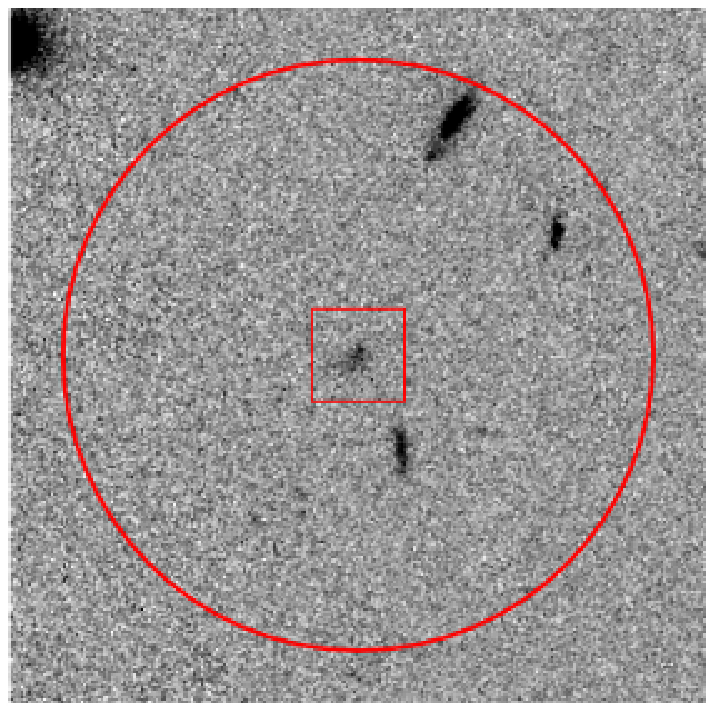}
\includegraphics[width=6cm]{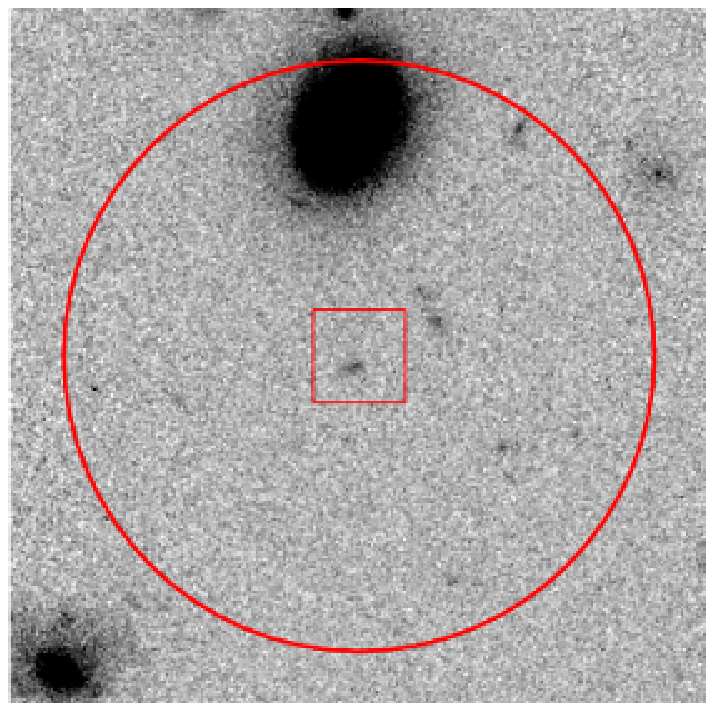}
\includegraphics[width=6cm]{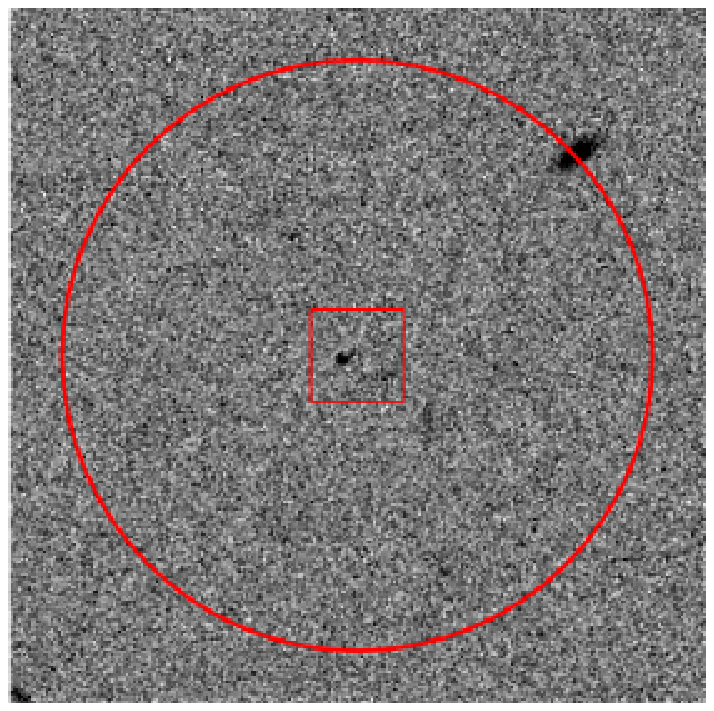}
\caption{HST-ACS images from the HST-COSMOS survey \citep{koekemoer07_HST}, 
centred on the six `free-line-of-sight'
background galaxies (see text for details). The red circle has a
radius of $6.4''$, corresponding to $50 \kpc$ (physical) at $z=2.9$. All
the other visible sources within the circle are spectroscopic galaxies
with $z_s < 2.8761$ or $z_s > 2.9133$ or photometric galaxies with
$z_p < 2.895-0.17$ or $z_p > 2.895+0.17$. } 
\label{iso_gals}
\end{figure*}

\begin{table*} 
\caption{Equivalent width (EW) measurements for the absorption feature at
$\lambda=4736$\AA, for different samples of background galaxies (all
background galaxies, background galaxies with redshift flag=3 and 4,
and `free-line-of-sight' background galaxies). The EW is rest frame 
and expressed in \AA. See Sect.~\ref{background} for details on the 
computation of the EW error and the definition of ``measurement S/N'' 
and ``detection S/N''. All these quantities are
computed for both the mean and median smoothed stacked spectrum.}
\label{EW_tab} 
\centering 
\begin{tabular}{l | r | r r r | r r r }
\hline\hline 
 {\bf Sample}    &  {\bf N$_{gal}$} & \multicolumn{3}{c|}{{\bf Mean}} & \multicolumn{3}{c}{{\bf Median}} \\    
  &                               & EW (r.f.) &  S/N meas. & S/N det.  &   EW (r.f.) &  S/N meas. & S/N det.\\    
\hline								   
Bkg galaxies, flag=2,3,4  &  36  & $4.0\pm1.2$	& 3.3 	   &	3.4    & $3.9\pm1.4$   &	2.9 &3.3    \\
Bkg galaxies, flag=3,4      &  18  & $5.1\pm1.5$ & 3.3	   &	3.9    & $6\pm1.8$     &	3.3 &4.0    \\
Bkg `free-line-of-sight'    &  6   & $10\pm2.2$  & 4.5	   &	5.9    & $10.8\pm3.7$  &	2.9 &3.9    \\
\hline 
\end{tabular} 
\end{table*}

\section{Searching for diffuse cold gas}\label{background}

In this section we describe our search for the presence of (diffuse)
cold gas within the proto-cluster. It has already been suggested that
galaxies could be fed by cold gas streams. The detection and study of
such gas, in the form of flows, blobs,  or diffuse nebulas, would add
precious pieces to the puzzle of galaxy evolution. Here, we try to
give constraints on the presence of this gas by examining the absorption
features in the spectra of galaxies in the background of our
structure, as already done, for example, in \cite{giavalisco11} for an
overdensity at $z=1.6$. We discuss our results in
Sect.~\ref{discussion}.

We inspected the spectra of the galaxies in the background of the
structure, searching for an absorption feature at the wavelength of
the Ly$\alpha$ at $z=2.895$, i.e.  $\lambda= 4736$\AA. Our aim is to
verify the presence of gas in the halo of the galaxies in the
overdensity or diffuse gas in the IGM of the overdensity itself.

These background galaxies are selected to be at $3\leq z\leq 4.15$,
with secure $z_s$ (flag=2, 3, 4, 9), and observed in the same VIMOS
quadrant as the structure. The lower limit in $z_s$ is required to
distinguish the possible absorption by Ly$\alpha$ at $z=2.895$ from
the intrinsic absorption of the Ly$\alpha$ in the given background
galaxy.  The upper $z_s$ limit excludes galaxies for which the line at
$\lambda= 4736$\AA~ (observed) falls blue-wards of the Lyman limit at
912\AA~ (rest frame). For such galaxies, we would not have signal in
the wavelength range of interest. With this selection, we found 36
background galaxies, amongst which 18, 12, and 6 with flag =2, 3, and 4
respectively. We also found one broad-line AGN with a secure redshift,
which we do not use in our analysis.

The top panel of Fig.~\ref{stacked_spec} shows the mean and median
stacked spectrum of all the 36 background galaxies  at observed
wavelengths. We compute the stacked spectrum in the following way.  We
interpolate each input spectrum on the same grid with a pixel scale of
5.3\AA/pixel (i.e. the dispersion of the blue grism used for the
observations). Even if the pixel scale is the same as in each single
spectrum, each of the spectra is interpolated on this pixel scale
before co-adding, and the flux rescaled to preserve the total flux
after rescaling. For each of the spectra, we compute the median sky
level, and we consider  `good pixels' to be those for which the sky flux
is within 120\% of the median sky level. In this way, in each
spectrum, we exclude the regions contaminated by
strong sky lines from the analysis. For each grid pixel at $\lambda_i$, the flux $f_i$
in the co-added spectrum is computed as the median (or mean) of the
fluxes at $\lambda_i$ of the input spectra, where only spectra for
which the pixel at $\lambda_i$ is a good pixel were considered.  For
each $\lambda_i$, we retain the information on the number of spectra
that have been used for the stacking ($N_{s,i}$). The stacked spectrum
is then smoothed on a scale comparable to the resolution of VUDS
spectra at the line of interest. Given R=230 (see Sect.~\ref{data}), 
we obtain a resolution element of $\Delta
\lambda \sim 25$\AA~ at $\lambda=4736$\AA. Spectra are neither normalised 
nor weighted before stacking.

In the spectra showed in Fig.~\ref{stacked_spec} we do see an absorption 
feature at $\lambda=
4736$\AA. In the median spectrum, this absorption feature has a rest
frame FWHM $\sim 9$\AA~ (not deconvolved with instrumental
resolution) and rest frame equivalent width $\EW \sim4 \pm 1.4$\AA, i.e. a 
measurement\footnote{We define ``measurement S/N'' the ratio
of the EW over its error, which indicates how well the EW has been
measured, while the ``detection S/N'' indicates how well the line is
overall detected above the continuum noise.} 
S/N of $\sim3$. Its detection S/N is $\sim3.3$. The error on the EW
has been computed in a similar way to \cite{tresse99}. For the mean
spectrum we obtain very similar values (see Table \ref{EW_tab}). If we
stack only the background galaxies with redshift flag=3 and 4 (mid
panel in Fig.~\ref{stacked_spec}), we find slightly larger EW in both
the mean and median spectra, but the values are compatible with the
ones obtained by stacking all galaxies (see Table \ref{EW_tab}).

Given that this feature is detected in both the mean and median
spectra with a reasonable S/N, this suggests that it is a common
feature in the sample of background galaxies and not due to a
minority of the spectra. This absorption feature is also visible (by
eye) in some of the single spectra. 
Clearly, this absorption could be caused by the presence of
an intervening galaxy in the structure and not to diffuse gas. For
instance, inspecting the HST image of the galaxy where the absorption
line at $\lambda= 4736$\AA~ is the most evident, we notice that there
are two faint objects close to the galaxy without any spectroscopic
or photometric redshift, one of the two possibly belonging to the
galaxy itself. In such a case
it is not possible to say if the absorption at $\lambda= 4736$\AA~ is
due to diffuse gas at $z\sim 2.895$ or to one of the two faint
objects, which could be at $z\sim2.895$.

To have a better handle on the possibility of detecting diffuse gas,
we removed this and similar cases from our background sample in the
following way. We looked in a radius of $50 \kpc$ (physical) around
each background galaxy and removed the given
background galaxy from our analysis if in that radius i) there was another VUDS galaxy
with $2.8761\leq z_s \leq 2.9133$; ii) there was a galaxy with $z_p$
(primary or secondary peak) in the range $z=2.895\pm0.17$; iii) there
was one or more unidentified sources (not included in our photometric
catalogue) in the HST image\footnote{ACS-HST images in the F814W filter 
are 50\% complete down to F814W$_{AB}$=26, see \cite{koekemoer07_HST}}. 
A possible absorption found in a background galaxy at
$\lambda= 4736$\AA, even if caused by gas at $z\sim 2.895$, would not be
resolved from an absorption line due to a foreground galaxy in the
range $2.8761\leq z_s \leq 2.9133$ (given the resolution of the
grism), and this consideration gives us the restriction on
spectroscopic galaxies assumed in point i). For galaxies
with only $z_p$ (point ii), we computed the distribution of $z_s- z_p$
for galaxies with $2.5<z_s<3.5$, fitted it with a Gaussian function
and found $\sigma\sim0.17$, which we used as the redshift interval
assumed in point ii).

After removing the background galaxies satisfying these three 
criteria, we are left with six `free-line-of-sight'
background galaxies. They are in the range $3.1 \lesssim z_s \lesssim
4.1$, and are distributed quite uniformly in $RA$-$Dec$, covering half
of the entire sky area of the structure (see bottom panel of
Fig.~\ref{radec}). Figure \ref{iso_gals} shows the HST images centred on
these six galaxies. Their median and mean coadded spectra  are shown in the
bottom panel of Fig.~\ref{stacked_spec}. In the median spectrum, the
absorption line at $\lambda= 4736$\AA~ has a flux detection S/N of
about 4 and a rest frame $\EW\sim11 \pm 4$\AA, i.e. an EW measurement
S/N of $\sim3$. The rest frame FWHM is $\sim 10$\AA.

In the stacked spectrum of the six `free-line-of-sight' galaxies there
are no other absorption features with a meaningful EW corresponding to
typical lines at $z=2.9$, but one. This exception is the (blended)
doublet Si IV $\lambda\lambda1393,1402$, with a rest frame EW of
$5.5\pm1.5$\AA.  This absorption feature is also visible in the two
other panels of Fig.~\ref{stacked_spec}, but with a lower EW
($\sim3\pm1$\AA). We also estimated the EW of the blends OI - Si II $\lambda\lambda1302,1304$,
and C II $\lambda1334$ in the coadded spectrum of galaxies with
redshift flag=3 and 4, and we find $\EW\sim3\pm1$ and
$\EW\sim2\pm1$\AA, respectively. Nevertheless, the lines are blended,
and they become even less evident when also coadding flag 2 and 9.

We verified that the absorption line at $\lambda= 4736$\AA~ in the
observed frame coadded spectrum of the six `free-line-of-sight' galaxies is not a
spurious effect of the noise of the coadded spectra. We did this by
comparing its EW with the EW of all the possible absorption features
(real or not) in 1000 stacked spectra, obtained by stacking six galaxies
chosen randomly amongst all the VUDS galaxies in the COSMOS field, with
redshift quality flag equal to 2, 3, 4, and 9, and with $z_s \geq 3$.
The result of this test does not change when the sets of six galaxies are built
considered only galaxies in the same redshift range as the six
galaxies with a free line of sight, i.e. $3.14 \leq z \leq 4.12$.

Appendix \ref{test_abs} describes in detail how we computed the
distribution of the observed-frame EWs of all the absorption features
in these 1000 stacked spectra. We find that the $25^{th}$, 
$50^{th}$, and $75^{th}$ percentile of this distribution
correspond to 3.6, 7.3, and 12.6\AA. The observed frame EW of the absorption line
in the median stacked spectrum of our six galaxies is 42\AA. This value
corresponds to the $99^{th}$ percentile of the distribution of all the
absorption features in the 1000 stacked spectra.
This result suggests that our measurement is unlikely to be
due to noise fluctuations. Moreover, in this analysis we did not make
any use of the additional information that the detected absorption
line at $\lambda=4736$\AA~ is exactly at the wavelength corresponding
to $Ly\alpha$ absorption at the redshift of the structure. This fact
reduces the likelihood that this line is spurious even further. The
physical interpretation of this significant feature is discussed
in Sect.~\ref{disc_gas}.

\section{Galaxy properties}\label{galaxies}

In this section we analyse the properties of the galaxies within the
overdensity, and we compare them with a control sample at the same
redshift but outside the overdensity.  For this analysis, we use 
only galaxies with redshift quality flag=3 and
4, so in the overdensity we retain only 11 galaxies out of 12 
(see Sect.~\ref{overdensity}). 
The control sample comprises all the VUDS galaxies with flag=3 and
4 in the range $2.8<z<3.0$, outside the structure. This sample
includes 151 galaxies. In the following analysis, we include 
all the galaxies in the two samples, irrespective of their 
stellar mass $\mathcal{M}$. We verified that the results do not change if 
we use only galaxies above a given mass limit, common to the 
two samples (i.e. ${\rm log}_{10}(\mathcal{M}/M_{\odot})\geq 9.5$).

A more detailed analysis is deferred to a future work, when the VUDS
selection function (Tasca et al, in prep.) will be fully assessed. A
robust computation of the selection function will allow us to quantify
the selection biases against specific population(s) due to the VUDS
observational strategy (such as the most massive and
passive galaxies, as described in Sect.~\ref{data_SED}). Nevertheless,
we note that here we perform a differential comparison of two
samples observed with the same strategy, and we do not attempt 
to compare our results here in an absolute way with other
samples from the literature.

We also stress that, given the small sample (11 galaxies in the
overdensity), the comparison of the galaxies in the overdensity with
any control sample is prone to large uncertainties. This kind of
analysis is better performed with larger samples of galaxies, but the
analysis presented here has the advantage of being one of the few ever 
attempted at these high redshifts.

\subsection{Stellar mass, star formation, and colours}\label{gal_SM}

Figure \ref{SM_SFR} shows the distribution of 
$\mathcal{M}$ and SFR for the two
samples. Both $\mathcal{M}$ and SFR are computed via SED fitting (see
Sect.~\ref{data_SED}). On the basis of a KS test, in both cases, the
two distributions are consistent with being drawn from the same
$\mathcal{M}$ and SFR distributions. As a complementary test, we
also verified the probability that a random sample of 11 galaxies,
extracted from the control sample, could lack the high-mass tail, as we
find in the proto-cluster (see the top panel of Fig.~\ref{SM_SFR}).
We randomly extracted 11 galaxies 1000 times from the control sample,
which has a maximum $\mathcal{M}$ of ${\rm
log}_{10}(\mathcal{M}/M_{\odot})\sim 11.5$. We found that, over the
1000 extractions, the maximum $\mathcal{M}$ has a mean value of ${\rm
log}_{10}(\mathcal{M}/M_{\odot})\sim 10.6$ with a dispersion of
0.35. This means that the maximum $\mathcal{M}$ found in the
proto-cluster (${\rm log}_{10}(\mathcal{M}/M_{\odot})\sim 10.2$) is
only $1.2 \sigma$ lower than the typical extracted one, confirming the
result of the KS test.

The top panel of Fig.~\ref{col_col} shows the $(NUV-r)$ vs $(r-K)$
rest frame colours. This plot is particularly useful for distinguishing
between active and passive galaxies, since it is able to distinguish dusty
active galaxies from truly red and passive ones (see
e.g. \citealp{williams09_UVJ,arnouts13_NUVrK}). Namely the $NUV-r$
colour is sensitive to specific star formation (sSFR: the higher the
sSFR, the bluer the colour), because $NUV$ traces recent star formation
and $r$ the old stellar populations. Dust attenuation can alter the
$NUV-r$ colour, moving dusty star-forming galaxies towards redder
colours. The $r-K$ colour does not vary much for different stellar
populations, but is very sensitive to dust attenuation (see
e.g. Fig.~3 in \citealp{arnouts13_NUVrK}). Using both colours, it is
possible to partially disentangle red passive galaxies from dusty
active ones. The top panel of Fig.~\ref{col_col} shows that possibly
the galaxies in the structure have a bluer $r-K$ colour (perhaps
meaning that are less attenuated by dust), but, given the lack of 
statistics, the difference in the $r-K$ distribution between the two
samples is not significant on the basis of a KS test. We note that
the locus of passive galaxies defined in \cite{arnouts13_NUVrK} is
$(NUV-r)>3.75$ for $(r-K)<0.4$ and $(NUV-r)>1.37 \times (r-K)+3.2$ for
$(r-K)\geq0.4$; i.e., it corresponds to $NUV-r$ colours that are much redder
than we find in our sample.

The bottom panel of Fig.~\ref{col_col} shows the sSFR versus the
stellar mass. Also in this case, the galaxies in the structure do not seem 
to have very different properties from those in the control
sample. At most, they could show a less scattered relation between
sSFR and $\mathcal{M}$, but given the error on the sSFR computation 
(see the cross in the top right corner of the panel), this difference 
is not significant.

We also analysed how galaxies are distributed in the plane $J-K$ vs
$K$ (observed magnitudes). The filters $J$ and $K$ bracket the
D4000\AA~ break at $z\sim3$, so the $J-K$ colour can be used to
identify passive galaxies (see e.g. \citealp{hatch11b}). We found that
galaxies within and outside the structure are distributed very
similarly in the $J-K$ vs $K$ plane. The reddest galaxies do not
belong to the overdensity, but to the field, and they do not seem to
reside primarily in the proximity of the overdensity. Considering
the two reddest spectroscopic galaxies in the structure, which have a
very similar $J-K$, one is located in the middle of P2Q1, the other
one close to its boundary.

In summary, we do not find any significant difference between the 
colours, stellar masses, SFR, and sSFR of the galaxies in the 
proto-cluster and in the control sample. While we cannot exclude 
that the lacking of any difference is real, it might also be due to the
lack of statistics, or to the fact that we miss, in our sample, a 
particular galaxy population that environment could affect more 
strongly (see e.g. the most passive galaxies, as discussed at the 
end of Sect.~\ref{data_SED}). A more careful analysis of this possible 
selection bias is deferred to a future work.

\begin{figure} \centering
\includegraphics[width=9cm]{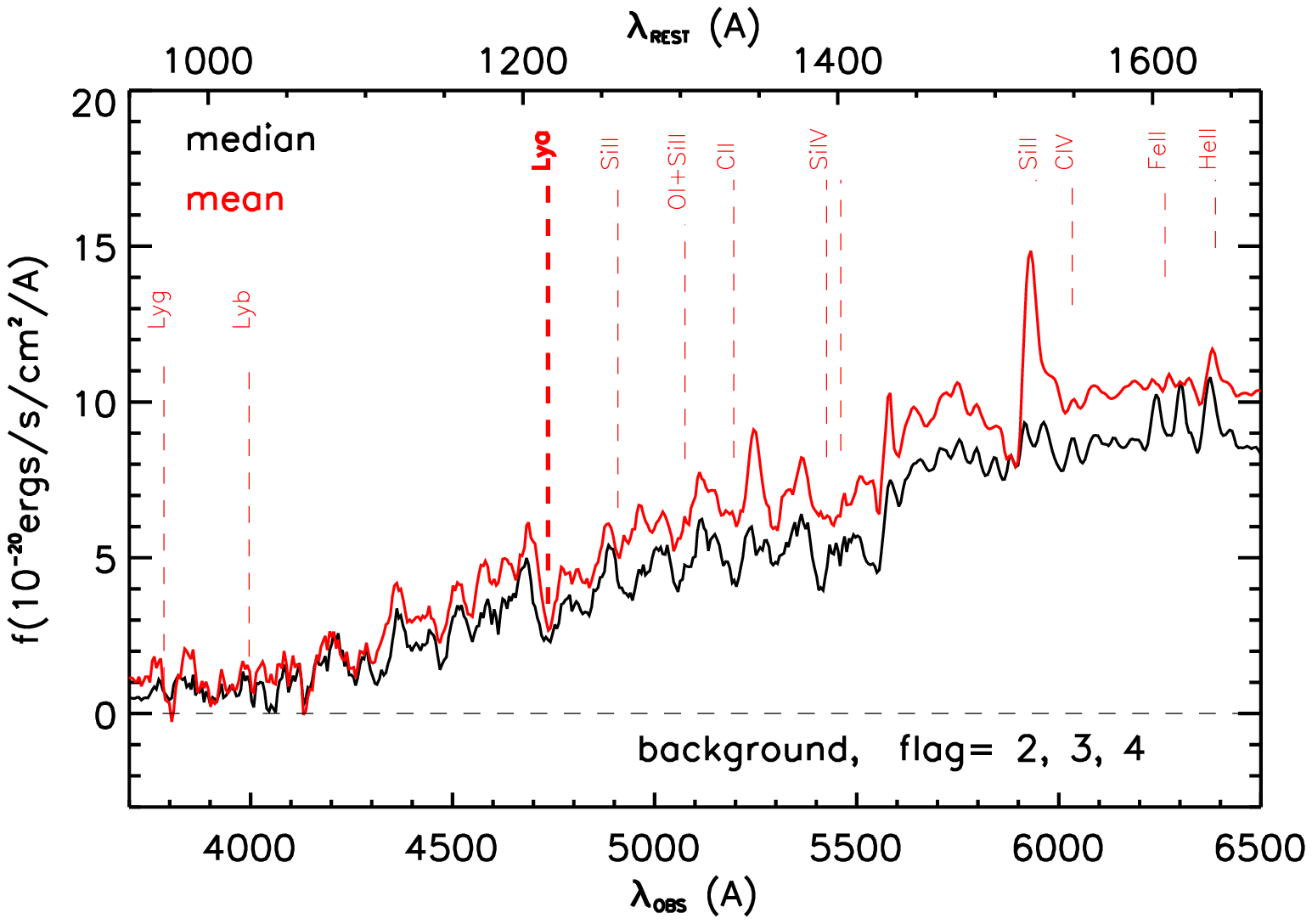}
\includegraphics[width=9cm]{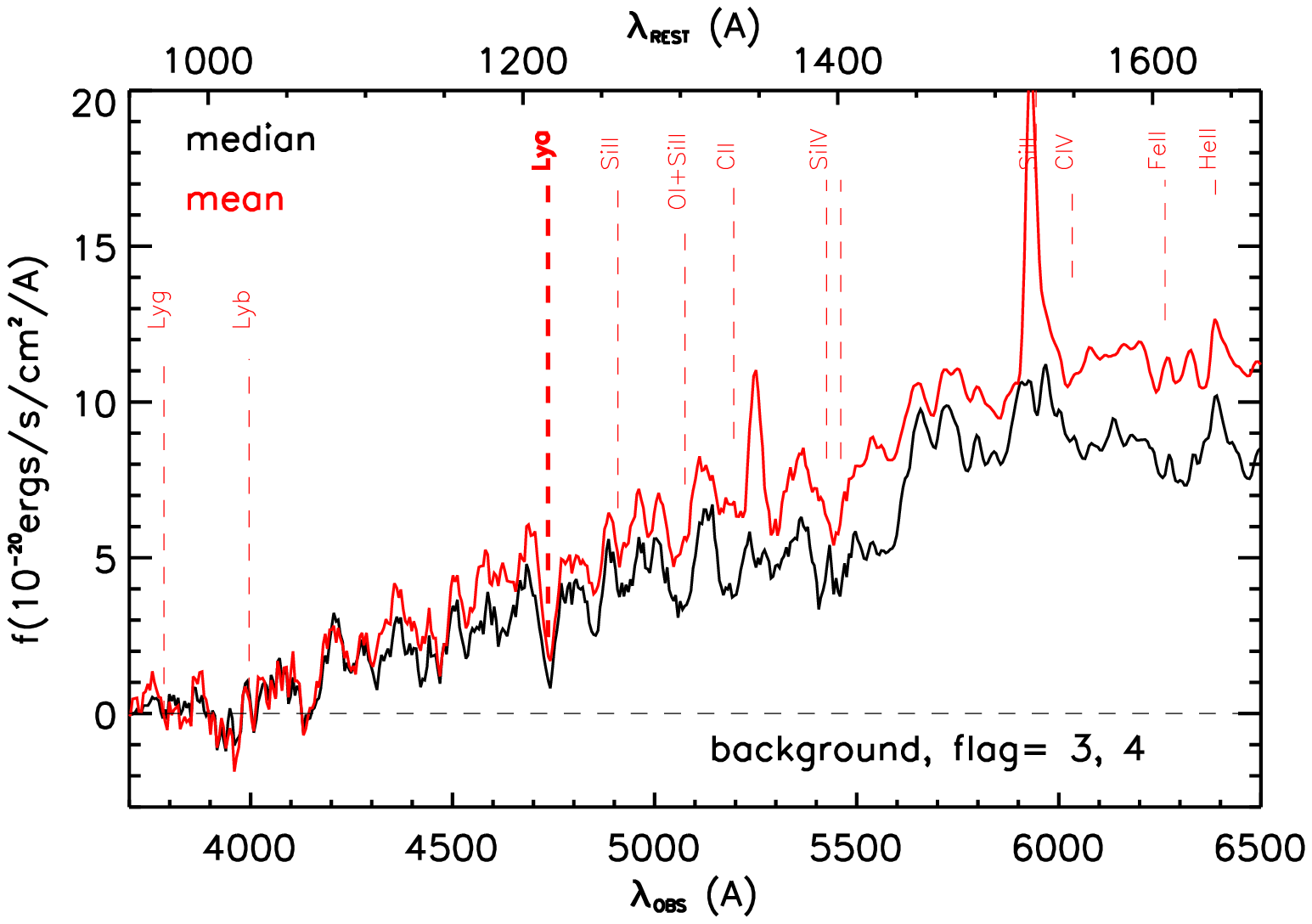}
\includegraphics[width=9cm]{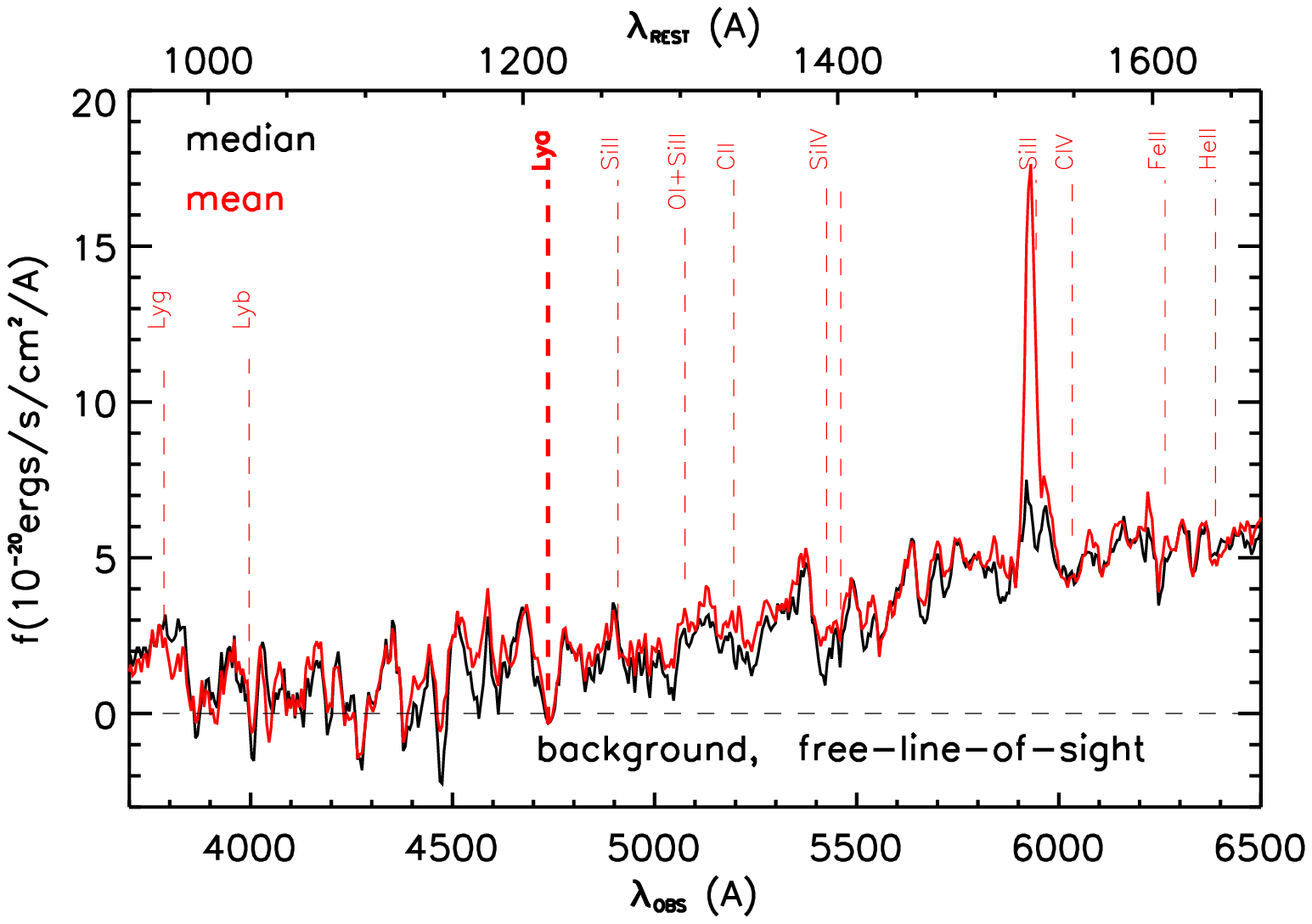}
\caption{{\it Top}. Median (black) and mean (red) stacked spectrum of all the background
galaxies at observed wavelengths. {\it Middle}. Same as top panel, but stacking only
background galaxies with redshift flag = 3 and 4.  {\it Bottom}. Same
as top panel, but stacking only the six `free line-of-sight'
background galaxies (see text for details). } 
\label{stacked_spec} \end{figure}

\subsection{Stacked spectra}\label{gal_spec}

Figure \ref{spec_in_out} shows the stacked spectra of the two
subsamples of galaxies, i.e. the 11 galaxies in the overdensity (top
panel) and the 151 galaxies in the control sample (bottom panel).  
Spectra are stacked as in Sect.~\ref{background}, but in this case 
they have been blue-shifted to rest frame wavelength before stacking,   
and the stacked spectrum is smoothed with a Gaussian filter with 
$\sigma$ equal to one pixel.

The two spectra show some differences. First, the median stacked
spectrum of the control sample shows some Ly$\alpha$ in emission,
indicating that at least half of the galaxies in the control sample
have Ly$\alpha$ in emission, while this is not the case for the median
spectrum of the galaxies in the overdensity. We verified that only 3
out of these 11 galaxies show an emission line at $\lambda=1216$\AA~
(rest frame). However, given the low statistics, this result does not
give us any clue to the typical Ly$\alpha$ in the galaxies in the
overdensity, given that the number of galaxies with Ly$\alpha$ in
emission in the structure is compatible, at $2\sigma$ level, with
being more than 50\%. In more detail, the three Ly$\alpha$
emitters in the proto-cluster have an EW(Ly$\alpha)> 25$\AA~ rest
frame, which makes them strong emitters according to the definition in
\cite{cassata14}. This means that the strong emitters are $27\pm 13$\%
of the proto-cluster sample. In the control sample, we find $19\pm3$\%
of strong Ly$\alpha$ emitters. The two fractions both
agree, within the error bars, with the overall percentage of
strong Ly$\alpha$ emitters at $z=3$ found in \cite{cassata14} for the
entire VUDS sample.

In the control sample coadded spectrum all the ISM absorption lines
are clearly visible: Si II $\lambda1260$, the blend OI - Si II
$\lambda\lambda1302,1304$, C II $\lambda1334$, the doublet Si IV
$\lambda\lambda1393,1402$, Si II $\lambda1526$, the doublet C IV
$\lambda\lambda1548,1550$, Fe II $\lambda1608$ and Al II
$\lambda1670$. Despite the higher noise, all these lines are also visible 
in the coadded spectrum of the overdensity galaxies, with the
exception of Fe II $\lambda1608$. The fact that we do not see a clear
Fe II $\lambda1608$ absorption is probably because  it
falls in a region where the observed frame spectrum is contaminated by
sky lines: the 11 galaxies in the structure are almost at the same
redshift, so the observed frame skylines in the 11 spectra always fall 
in the same rest-frame wavelength. This does not happen for the
galaxies in the control sample, which span a larger extent in redshift.
All the absorption lines in the proto-cluster coadded spectrum
have EWs that are compatible, within the error bars, with those in the
coadded spectrum of the control sample.

\begin{figure} \centering
\includegraphics[width=8cm]{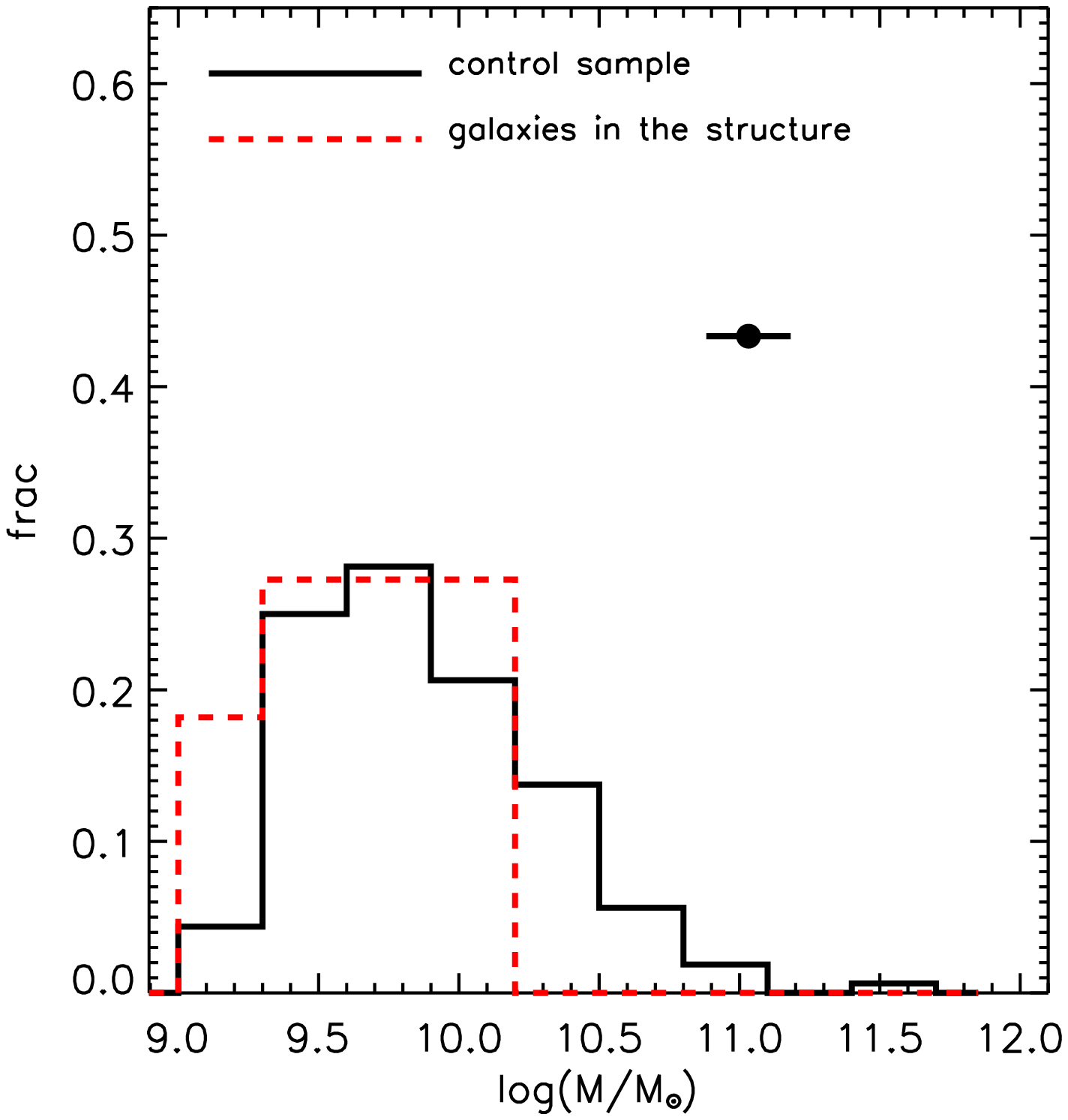}
\includegraphics[width=8cm]{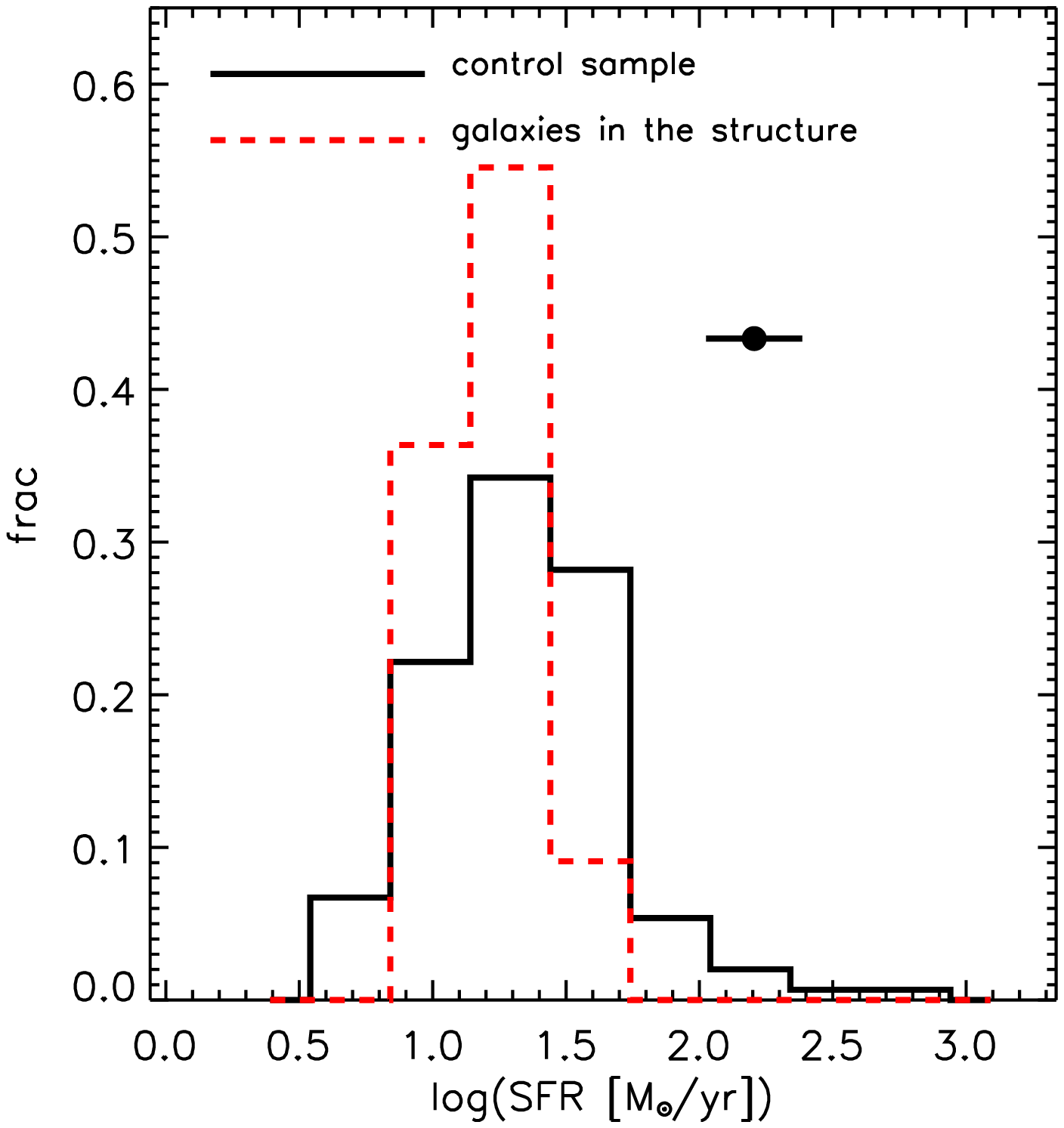}
\caption{{\it Top.} Distribution of galaxy stellar masses in the control 
sample (black solid line) and in the structure (red dashed line). 
{\it Bottom.} Same as in the top panel,
but for the distribution of SFR. In both panels, the solid point with the error bar shows the 
typical error on $\mathcal{M}$ and SFR. The possible systematic offset in the SFR 
computation due to the choice of different SFHs (see Sect.~\ref{data_SED}) is not 
included.} 
\label{SM_SFR} \end{figure}

\begin{figure} \centering
\includegraphics[width=8cm]{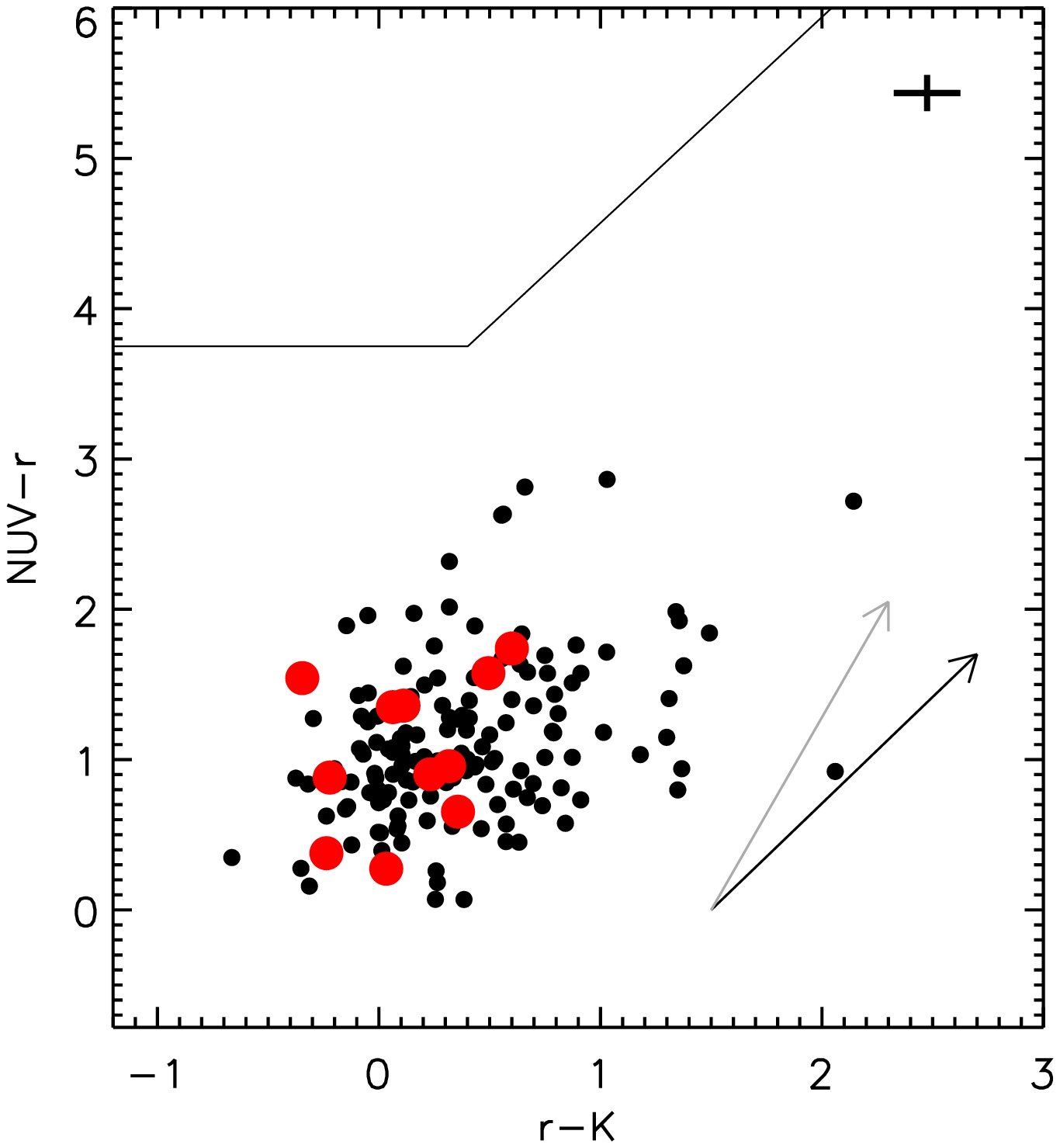}
\includegraphics[width=8cm]{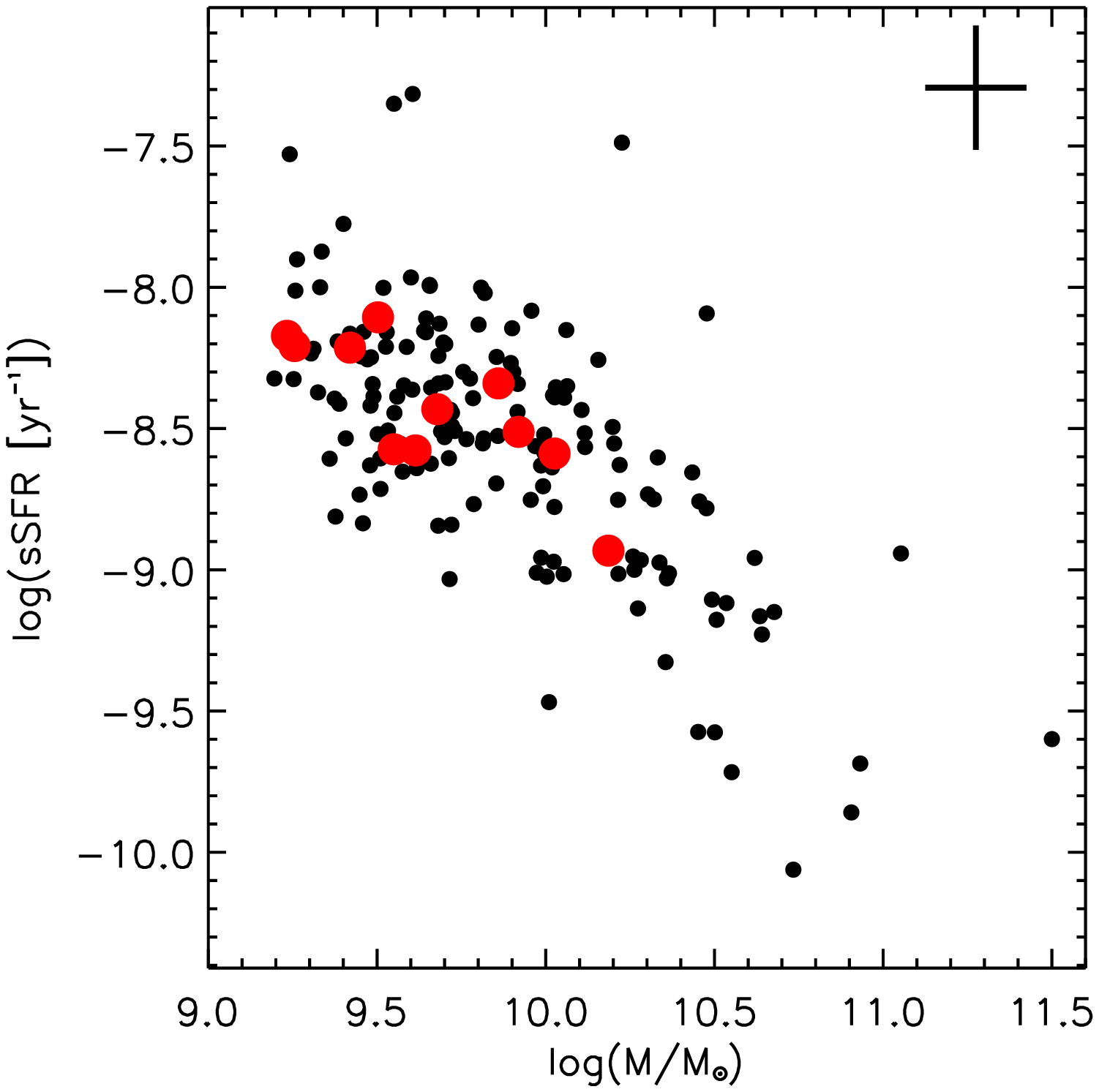}
\caption{{\it Top.} $NUV-r$
vs $r-K$ rest frame colours. Black dots are the control sample,
red filled circles are the galaxies in the structure. The solid thick lines 
comprise the locus of passive 
galaxies defined in \cite{arnouts13_NUVrK}. The two arrows represent
the attenuation vectors for starburst (dark arrow) 
and Small Magellanic Cloud (light arrow) attenuation curves 
assuming $E(B-V) = 0.4$ as in  \cite{arnouts13_NUVrK} (see their Fig.2). 
{\it Bottom.} sSFR versus stellar mass. Symbols as in the top panel. 
In both panels, the cross in the top right corner shows the 
typical error on the quantities plotted on the two axes. The possible 
systematic offset in the SFR computation due to the choice of different 
SFHs (see Sect.~\ref{data_SED}) is not included when deriving the error on the 
sSFR.} 
\label{col_col} \end{figure}

\begin{figure} \centering
\includegraphics[width=8cm]{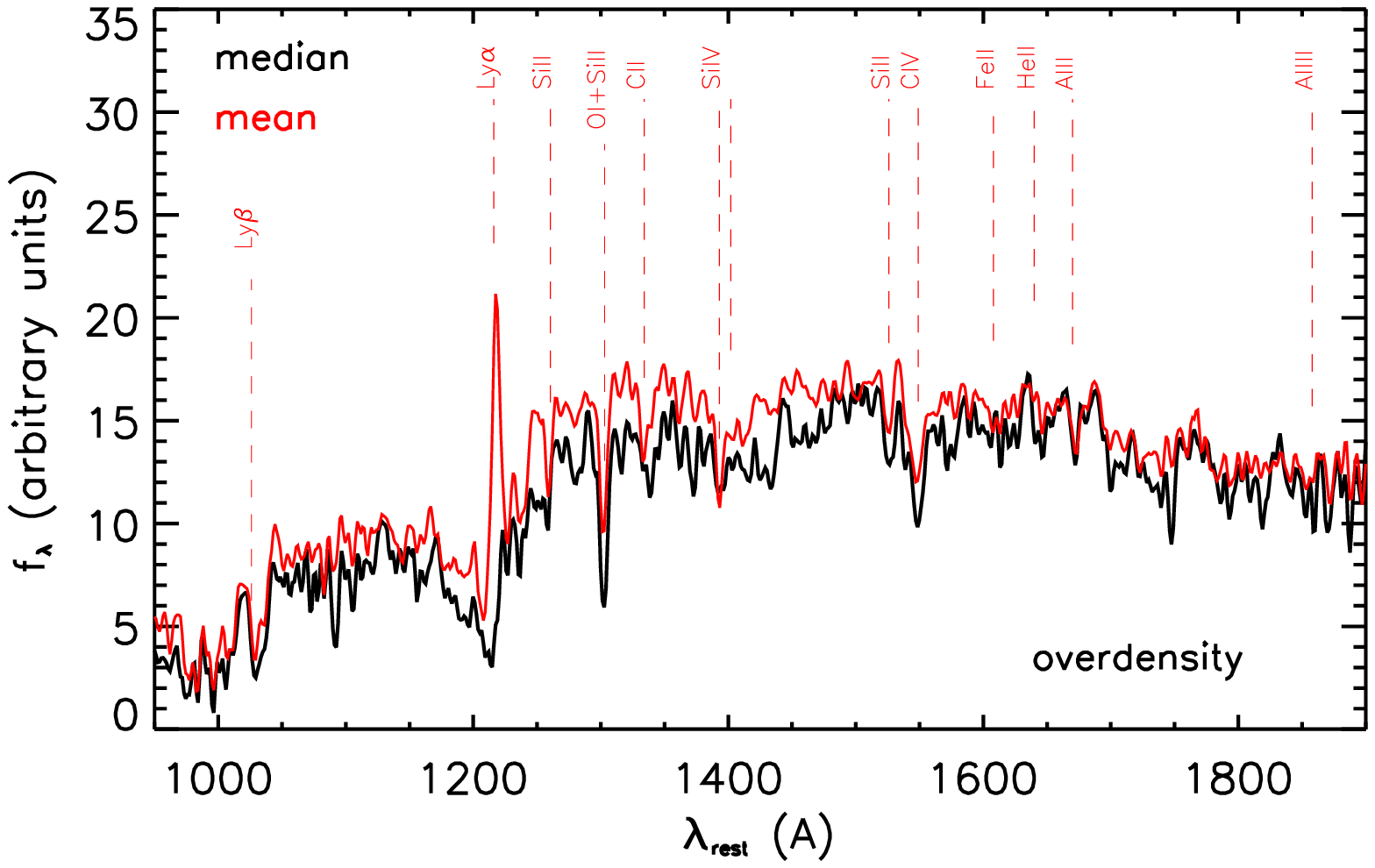}
\includegraphics[width=8cm]{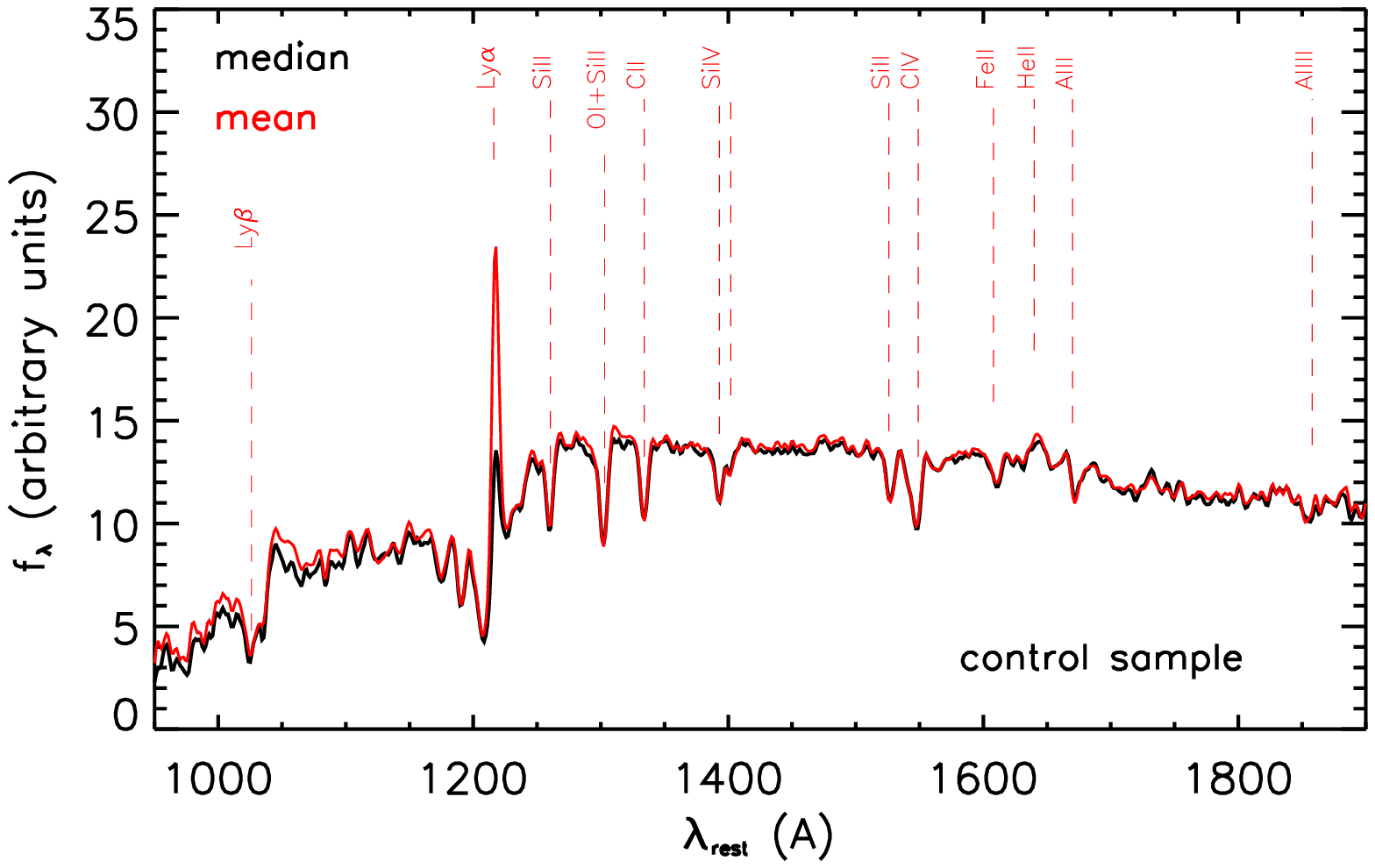}
\caption{{\it Top:} coadded spectrum of the 11 galaxies in the
structure with redshift quality flags 3 and 4. {\it Bottom:} coadded
spectrum of the 151 galaxies in the control sample with redshift
quality flags 3 and 4. Line and colour codes are as in Fig.~\ref{stacked_spec}.} 
\label{spec_in_out} \end{figure}


\section{Discussion}\label{discussion}

The overdensity that we found at $z\sim2.9$ in the VUDS sample in the
COSMOS field has two main characteristics that make it an extremely
interesting case study: the high value of the galaxy overdensity 
and the evidence of the presence of cold gas in the IGM. In this
section we discuss in more details these two points.

\subsection{The overdensity}\label{disc_delta}

In Sect.~\ref{overdensity}, we estimated a galaxy overdensity of
$\delta_g =12 \pm 2$. This overdensity is much higher than the
typical $\delta_g$ found in the literature at similar redshift (e.g. 
 $\delta_g \sim 6$ at $z = 3.09$ in \citealp{steidel00}, $\delta_g
=2.3$ at $z=3.13$ in \citealp{venemans07}; see also Table 5 in
\citealp{chiang13_sim} for a more complete list, and their more recent
work shown in \citealp{chiang14_cosmos}\footnote{One of the
proto-cluster candidates found in \cite{chiang14_cosmos} is very
close in RA, Dec and redshift to our overdensity, as we discuss in
Sect.~\ref{overdensity}, but it has been detected by
\cite{chiang14_cosmos} with a galaxy density contrast much lower
than ours, namely $\delta_{g}=1.35$.}). Overdensities with $\delta_g$
similar to ours have been found, but they are very rare, and they seem
to span a wide redshift range (e.g. $\delta_g \sim 8.3$ at $z = 1.6$
in \citealp{kurk09}, $\delta_g\sim16$ at $z\sim6$ in
\citealp{toshikawa12}). We also refer the reader to \cite{lemaux14},
where we study a spectroscopic overdensity of $\delta_g=13.3\pm6.6$
found at $z=3.3$ in the VVDS field using VUDS observations.

The study of such over-dense regions is particularly important because 
these structures are more likely to become clusters than are lower density
structures \citep{chiang13_sim}. This allows us to study how mass
assembles along cosmic time. In our case, the high $\delta_g$ value is
primarily due to the narrow $z$ range ($\Delta z = 0.016$) in which the
spectroscopic members reside compared with much larger redshift slices
in which other overdensities have been found.

The large $\delta_g$ means a high total mass (given the typical bias
between galaxies and matter at $z=2.9$). Specifically, we find that
the total mass associated to the structure at $z=2.9$ is $\sim (8.1
\pm 2.4) \times 10^{14} M_{\sun}$, which makes our overdensity one of
the most massive found in the literature at high redshift. Following
\cite{chiang13_sim}, we estimate that this structure will collapse in
a cluster with $M_{z=0}\sim(2.5\pm0.8) \times 10^{15} M_{\odot}$ at
$z=0$. For both masses, the uncertainty represents $\sim30$\% of the
nominal value (see Sect.~\ref{mass}). We verify how rare these
structures might be and how many we should expect to find in
the volume explored by VUDS in the COSMOS field. Considering the snapshot at 
$z=0$ of the Millennium Simulation\footnote{We do not use 
our 10 light cones for this test, since in these light cones we do not find any structure
at $z\sim2.9$ that will collapse at $z=0$ in a cluster with
$M_{z=0}\sim(2.5\pm0.8) \times 10^{15} M_{\odot}$}, we find a density of $(1.1\pm
0.5)\times 10^{-8}$ clusters/Mpc$^3$ for clusters with
$M_{z=0}\sim(2.5\pm0.8) \times 10^{15} M_{\odot}$. This result is in
rough agreement with the cluster mass function at $z\sim0$ in
\cite{vikhlinin09} (see their Fig.~1): in their work, the density of
clusters with mass $M_{z=0}\sim(2.5\pm0.8) \times 10^{15} M_{\odot}$
(converting their M$_{500}$ in M$_{200}$, to be consistent with the
mass in the Millennium Data Base) is
$\sim(2.5\pm0.9\times10^{-8}$Mpc$^{-3}$. Given this density range
($\sim(1.1-2.5)\times10^{-8}$Mpc$^{-3}$), we expect to find 0.1-0.3
proto-clusters with such $M_{z=0}$ in the volume covered by VUDS in
the COSMOS field in the entire redshift extent $2<z<3.5$.

At face value, the measurement of a total mass of $\sim 8 \times
10^{14} M_{\sun}$ at $z\sim2.9$ and a total mass of $\sim 2.5 \times
10^{15} M_{\sun}$ at $z=0$ seems to suggest a very small evolution in
mass. This is not true, since the two masses are not directly
comparable. The mass measured at $z=0$ is the mass bound in the
cluster, while at $z\sim2.9$ we measure the total mass that will
collapse in that cluster and that at $z\sim2.9$ is not necessarily
bound in one single structure. The two masses should differ only by
the factor $C_e$ described by \cite{chiang13_sim}, which links the
total mass at $z\sim2.9$ enclosed in a given volume with the final
bound mass at $z=0$. The most evident evolution from the overdensity at
$z\sim2.9$ and its descendant cluster at $z=0$ is an evolution in
overdensity and not in total mass.

A different discussion has to be devoted to the measured velocity
dispersion. We measured $\sigma_{los}=270\pm 80 \kms$ with the
biweight method, a method that is quite effective in measuring a
dispersion of a contaminated Gaussian distribution. Indeed, it is not
known which should be the shape of the velocity distribution of
galaxies in a proto-cluster, if such structure is not already relaxed. 
We inspected the shape of the velocity distribution of the galaxies in
the proto-clusters at $z=2.9$ in the galaxy mock catalogues that we
used in Sect.~\ref{mill}, focusing our attention on the proto-clusters
that will collapse in clusters with $M_{z=0}\gtrsim 10^{15}
M_{\odot}$. We found that these velocity distributions can have a
broad range of shapes, from Gaussian, to contaminated Gaussian, to
almost flat distributions. Indeed, if our measured $\sigma_{los}$ is
lower than the expected one (see \citealp{eke98}), there are cases in
the literature where the measured $\sigma_{los}$ is too large with
respect to expectations (see e.g. \citealp{toshikawa12}). We suggest
that the important piece of information that we can retrieve from
simulations is the maximum span in redshift space that the member
galaxies can encompass, rather than their velocity dispersion. The
first one is needed to tune the search for high-$z$ overdensities,
while the second one might be less meaningful, depending on the
evolutionary state of the proto-cluster under analysis.

\subsection{The diffuse gas}\label{disc_gas}

The stacked spectrum of the background galaxies at $z > 3.0$ shows an
absorption consistent with Ly$\alpha$ at z = 2.895 ($\sim 4736
\AA$). Its rest frame EW is $\sim 4-10$\AA~ according to the sample
used for the stacking (see Table~\ref{EW_tab}). These EWs correspond
to N(HI) $\sim 3-20\times 10^{19}$cm$^{-2 }$ (with large
uncertainties and assuming a constant gas density over the transverse
dimension of the structure). The angular size of the structure would
imply a total mass of HI of a few $10^{12}-10^{13} M_{\odot}$ if the
absorption were due to diffuse gas.

Several hypotheses can be suggested on the very nature of this
absorption feature. The debate is open in the literature about the
detectability of the signature of physical processes that could give
rise to such a feature. The first hypothesis, already introduced above
for the computation for the total HI mass, is that the Ly$\alpha$ in
absorption is related to the diffuse gas (IGM) within the
overdensity. This could be the case especially for the six
`free-line-of-sight' galaxies, because they are $>50 \kpc$ (physical) away
from all the identified sources in the structure.

The high column density that we infer could support a second
hypothesis that the absorption is due to intervening cold streams of
gas, which are falling into the potential well of the galaxies in the
overdensity. This high column density is shown, for instance, in
\cite{goerdt12}. They use hydrodynamical simulations at $z\sim3$ to
create and trace such streams in the circumgalactic medium via
Ly$\alpha$ absorption manifest in the spectra of background
galaxies. In particular, their Fig.~15 shows that the projected HI
column density of such absorption features can reach up to a few
$10^{20}$cm$^{-2}$ at a distance of more than $50 \kpc$ from the
galaxy that is fed by cold streams.  This N(HI) value is consistent
with the N(HI) of the absorption line that we find in the stacked
spectrum of the `free-line-of-sight' background galaxies. Similarly,
they show in their Fig.~20 that the Ly$\alpha$ rest frame EW can
easily reach values of $\EW\sim4-5$\AA~ along the line of sight of the
cold flows up to $60-80\kpc$ from the centre of the galaxy.

The fraction of the area (or `covering factor') around a galaxy that
is covered by these cold streams with high enough density to produce a
detectable absorption line is very low. For $\EW({\rm
Ly}\alpha)\sim4-5$\AA, \cite{goerdt12} show that the covering factor
is typically a few percent. Given these results for the
covering factor, the fact that we see such a strong absorption not
only in the mean stacked spectrum, but also in the median, strongly
suggests that the absorption is due to a more widely extended source
rather than to cold streams. On the other hand, the fact that we are
studying an over-dense region would likely give rise to denser and/or
more widely distributed filaments (see e.g. \citealp{kimm11}, who find
that the covering fraction of filaments is larger in more massive
haloes).

Very similar results for simulated cold streams extension, N(HI), and
covering factor are shown in \cite{fumagalli11} (see e.g. their
Fig~6). They also show that such cold flows are generally metal poor
(as also shown in \citealp{goerdt12} and \citealp{kimm11}). They suggest
that any observational detection of metal lines is more probably
due to outflows rather than infalls. We do not find any other
significant absorption lines in the coadded spectrum of the
`free-line-of-sight' background galaxies except for the Si IV
$\lambda\lambda1393,1402$ (blended) doublet, with a rest frame EW of
$5.5\pm1.5$\AA~ (EW $=3\pm1$\AA~ for the total sample of background
galaxies). Such an EW could, however, be because this is
a not resolved doublet. This detection, which does not directly support  
the infall scenario, will deserve further analysis.

\section{Summary and conclusions}\label{summary}

In this work, we have characterised a massive proto-cluster at $z=2.895$
that we found in the COSMOS field using the spectroscopic sample of
the VUDS survey. Our results can be summarised as follows:

\begin{itemize}

\item[-] The overdensity comprises 12 galaxies with secure
spectroscopic redshift in an area of $\sim7'\times8'$, in a total
redshift range of $\Delta z = 0.016$. The measured galaxy overdensity
is $\delta_g =12 \pm 2$. According to simulations \citep{chiang13_sim}, a structure
with $\delta_g =12$ at $z\sim2.9$ has a 100\% probability of evolving into
a galaxy cluster at $z=0$.

\item[-] We estimated that this overdensity has a total mass of
$M\sim8.1\times10^{14}M_{\odot}$ in a volume $13\times15\times17$
Mpc$^3$. According to \cite{chiang13_sim}, such an 
overdensity should collapse into a cluster of total mass
$M_{z=0}\sim2.5 \times 10^{15} M_{\odot}$ at $z=0$. In the volume
surveyed by VUDS at $2<z<3.5$ in the COSMOS field, we should have
expected 0.12-0.3 proto-cluster of this kind.

\item[-] The velocity dispersion of the 12 members is
$\sigma_{los}=270 \pm 80 \kms$. We used light cones
extracted from the Millennium Simulation to verify that this is lower (but
consistent within $2\sigma$) than the typical velocity dispersion of
the galaxies belonging to the same kind of proto-clusters at this
redshift. This low value is consistent with the increase in
$\sigma_{los}$ as time goes by \citep{eke98}.

\item[-] In the light cones that we examined, the typical span in
redshift of the galaxies belonging to proto-clusters at $z\sim2.9$,
which will collapse into massive clusters at $z=0$, is $\Delta z \sim
0.02$. This value is much lower than the redshift bin often used to
search for proto-clusters at this redshift.

\item[-] The stacked spectra of the galaxies in the background of
the overdensity show a significant absorption feature at the observed
wavelength corresponding to the Ly$\alpha$ at the redshift of the
structure ($\lambda=4736$\AA). We find that this absorption feature
has a rest frame EW of $10.8\pm3.7$\AA, with a detection S/N of
$\sim4$, when stacking only background galaxies without intervening
sources at $z\sim2.9$ along their line of sight.  We verified that
this measurement is likely not to be due to noise fluctuations. Considering
also the lower (but consistent) EW found using different samples of
background galaxies (see Table \ref{EW_tab}), such an EW range corresponds to
a column density N(HI) of the order of $3-20\times10^{19}$cm$^{-2}$.

\item[-] We analysed the properties of the galaxies within the
overdensity, and we compared them with a control sample at
approximately the same redshift outside the overdensity.  We
could not find any statistically significant difference between the
properties (stellar mass, SFR, sSFR, $NUV-r$, $r-K$) of the galaxies
inside and outside the overdensity, but this result might be due to
the lack of statistics, or possibly to the specific galaxy population
sampled by VUDS, which could be less affected by environment than other
populations not probed by the survey.

\end{itemize}

Simulations \citep{chiang13_sim} indicate that such an overdensity at
$z\sim2.9$ is indeed a proto-cluster that, given the measured galaxy
overdensity, will collapse to a (massive) cluster at $z=0$.  For this
reason, the detailed analysis of this proto-cluster represents a
fundamental step in the comprehension of galaxy formation and
evolution.

For the properties of the galaxies within the
proto-cluster, we plan to do a more detailed study when the VUDS
selection function is assessed. A well-defined selection function will
allow us to robustly quantify the average properties of the galaxies
in such a dense environment and to compare them with the galaxies in
the field at the same $z$ and with other overdensities found in the
literature (e.g. \citealp{lemaux14}). 
The synergy of spectroscopy and multi-band photometry in next-generation
surveys like Euclid will allow several proto-cluster
structures to be identified thanks to combination of depth and large surveyed areas.
In this respect, current surveys such as VUDS, are essential for 
characterising the properties of these structures and using
them to predict their observability and optimal detection with
future surveys.

On the side of the detection of cold gas, the EW of the absorption
line corresponding to the Ly$\alpha$ at $z\sim2.9$ implies a high
column density (N(HI)$\sim10^{20}$cm$^{-2}$). This N(HI) value would
be compatible with the scenario where the absorption is due to 
intervening cold
streams of gas, which are falling (and feeding) into the halo's potential
well of the galaxies in the proto-cluster
\citep{fumagalli11,goerdt12}. In contrast, the stacked spectrum of 
the galaxies  in the proto-cluster background also shows an absorption 
line corresponding to Si IV
$\lambda\lambda1393,1402$ (blended) doublet at the redshift of the 
proto-cluster, detection that does not agree with the prediction 
that the cold flows are  metal poor.
Surely, the scenario of gas accretion by cold gas stream needs to be more robustly
assessed from additional 
observational evidence. An exciting prospect is to look for the Ly$\alpha$ 
emission produced by gravitational energy released by cold gas 
flowing into the potential wells of galaxies (see e.g. \citealp{goerdt10}).


\begin{acknowledgements} 

We thank the referee for the useful comments that improved the paper.
We thank Nico Cappelluti for his help with X-ray data. We thank ESO
staff for their continuous support of the VUDS survey, particularly
the Paranal staff conducting the observations and Marina Rejkuba and
the ESO user support group in Garching.  This work is supported by
funding from the European Research Council Advanced Grant
ERC-2010-AdG-268107-EARLY and by INAF Grants PRIN 2010, PRIN 2012, and
PICS 2013. AC, OC, MT, and VS acknowledge the grants MIUR PRIN
2010--2011 and ASI n.I/023/12/0 ``Attivit\`a relative alla fase B2/C
per la missione Euclid''.  DM gratefully acknowledges LAM hospitality
during the initial phases of the project.  This work is based on data
products made available at the CESAM data centre, Laboratoire
d'Astrophysique de Marseille. This work partly uses observations
obtained with MegaPrime/MegaCam, a joint project of the CFHT and
CEA/DAPNIA, at the Canada-France-Hawaii Telescope (CFHT), which is
operated by the National Research Council (NRC) of Canada, the
Institut National des Sciences de l'Univers of the Centre National de
la Recherche Scientifique (CNRS) of France, and the University of
Hawaii. This work is based in part on data products produced at
TERAPIX and the Canadian Astronomy Data Centre as part of the
Canada-France-Hawaii Telescope Legacy Survey, a collaborative project
of the NRC and CNRS.

\end{acknowledgements}

\bibliographystyle{aa}
\bibliography{biblio}

\begin{thebibliography}{83}
\expandafter\ifx\csname natexlab\endcsname\relax\def\natexlab#1{#1}\fi

\bibitem[{{Andreon} {et~al.}(2009){Andreon}, {Maughan}, {Trinchieri}, \&
  {Kurk}}]{andreon09}
{Andreon}, S., {Maughan}, B., {Trinchieri}, G., \& {Kurk}, J. 2009, \aap, 507,
  147

\bibitem[{{Arnouts} {et~al.}(1999){Arnouts}, {Cristiani}, {Moscardini},
  {Matarrese}, {Lucchin}, {Fontana}, \& {Giallongo}}]{arnouts99}
{Arnouts}, S., {Cristiani}, S., {Moscardini}, L., {et~al.} 1999, \mnras, 310,
  540

\bibitem[{{Arnouts} {et~al.}(2013){Arnouts}, {Le Floc'h}, {Chevallard},
  {Johnson}, {Ilbert}, {Treyer}, {Aussel}, {Capak}, {Sanders}, {Scoville},
  {McCracken}, {Milliard}, {Pozzetti}, \& {Salvato}}]{arnouts13_NUVrK}
{Arnouts}, S., {Le Floc'h}, E., {Chevallard}, J., {et~al.} 2013, \aap, 558, A67

\bibitem[{{Bardelli} {et~al.}(1998){Bardelli}, {Pisani}, {Ramella}, {Zucca}, \&
  {Zamorani}}]{bardelli98}
{Bardelli}, S., {Pisani}, A., {Ramella}, M., {Zucca}, E., \& {Zamorani}, G.
  1998, \mnras, 300, 589

\bibitem[{{Beers} {et~al.}(1990){Beers}, {Flynn}, \& {Gebhardt}}]{beers1990}
{Beers}, T.~C., {Flynn}, K., \& {Gebhardt}, K. 1990, \aj, 100, 32

\bibitem[{{Bielby} {et~al.}(2013){Bielby}, {Hill}, {Shanks}, {Crighton},
  {Infante}, {Bornancini}, {Francke}, {H{\'e}raudeau}, {Lambas}, {Metcalfe},
  {Minniti}, {Padilla}, {Theuns}, {Tummuangpak}, \& {Weilbacher}}]{bielby13}
{Bielby}, R., {Hill}, M.~D., {Shanks}, T., {et~al.} 2013, \mnras, 430, 425

\bibitem[{{Borgani} {et~al.}(1999){Borgani}, {Girardi}, {Carlberg}, {Yee}, \&
  {Ellingson}}]{borgani1999}
{Borgani}, S., {Girardi}, M., {Carlberg}, R.~G., {Yee}, H.~K.~C., \&
  {Ellingson}, E. 1999, \apj, 527, 561

\bibitem[{{Bouch{\'e}} {et~al.}(2013){Bouch{\'e}}, {Murphy}, {Kacprzak},
  {P{\'e}roux}, {Contini}, {Martin}, \& {Dessauges-Zavadsky}}]{bouche13}
{Bouch{\'e}}, N., {Murphy}, M.~T., {Kacprzak}, G.~G., {et~al.} 2013, Science,
  341, 50

\bibitem[{{Bruzual} \& {Charlot}(2003)}]{BC03}
{Bruzual}, G. \& {Charlot}, S. 2003, \mnras, 344, 1000

\bibitem[{{Calzetti} {et~al.}(2000){Calzetti}, {Armus}, {Bohlin}, {Kinney},
  {Koornneef}, \& {Storchi-Bergmann}}]{calzetti2000}
{Calzetti}, D., {Armus}, L., {Bohlin}, R.~C., {et~al.} 2000, \apj, 533, 682

\bibitem[{{Capak} {et~al.}(2007){Capak}, {Aussel}, {Ajiki}, {McCracken},
  {Mobasher}, {Scoville}, {Shopbell}, {Taniguchi}, {Thompson}, {Tribiano},
  {Sasaki}, {Blain}, {Brusa}, {Carilli}, {Comastri}, {Carollo}, {Cassata},
  {Colbert}, {Ellis}, {Elvis}, {Giavalisco}, {Green}, {Guzzo}, {Hasinger},
  {Ilbert}, {Impey}, {Jahnke}, {Kartaltepe}, {Kneib}, {Koda}, {Koekemoer},
  {Komiyama}, {Leauthaud}, {Lefevre}, {Lilly}, {Liu}, {Massey}, {Miyazaki},
  {Murayama}, {Nagao}, {Peacock}, {Pickles}, {Porciani}, {Renzini}, {Rhodes},
  {Rich}, {Salvato}, {Sanders}, {Scarlata}, {Schiminovich}, {Schinnerer},
  {Scodeggio}, {Sheth}, {Shioya}, {Tasca}, {Taylor}, {Yan}, \&
  {Zamorani}}]{capak2007_photcat}
{Capak}, P., {Aussel}, H., {Ajiki}, M., {et~al.} 2007, \apjs, 172, 99

\bibitem[{{Capak} {et~al.}(2011){Capak}, {Riechers}, {Scoville}, {Carilli},
  {Cox}, {Neri}, {Robertson}, {Salvato}, {Schinnerer}, {Yan}, {Wilson}, {Yun},
  {Civano}, {Elvis}, {Karim}, {Mobasher}, \& {Staguhn}}]{capak11}
{Capak}, P.~L., {Riechers}, D., {Scoville}, N.~Z., {et~al.} 2011, \nat, 470,
  233

\bibitem[{{Cappelluti} {et~al.}(2007){Cappelluti}, {Hasinger}, {Brusa},
  {Comastri}, {Zamorani}, {B{\"o}hringer}, {Brunner}, {Civano}, {Finoguenov},
  {Fiore}, {Gilli}, {Griffiths}, {Mainieri}, {Matute}, {Miyaji}, \&
  {Silverman}}]{cappelluti07_XMM}
{Cappelluti}, N., {Hasinger}, G., {Brusa}, M., {et~al.} 2007, \apjs, 172, 341

\bibitem[{{Cassata} {et~al.}(2014){Cassata}, {Tasca}, {Le Fevre}, {Lemaux},
  {Garilli}, {Le Brun}, {Maccagni}, {Pentericci}, {Thomas}, {Vanzella},
  {Zamorani}, {Zucca}, {Amorin}, {Bardelli}, {Capak}, {Cassara'}, {Castellano},
  {Cimatti}, {Cuby}, {Cucciati}, {de la Torre}, {Durkalec}, {Fontana},
  {Giavalisco}, {Grazian}, {Hathi}, {Ilbert}, {Moreau}, {Paltani}, {Ribeiro},
  {Salvato}, {Schaerer}, {Scodeggio}, {Sommariva}, {Talia}, {Taniguchi},
  {Tresse}, {Vergani}, {Wang}, {Charlot}, {Contini}, {Fotopoulou},
  {Lopez-Sanjuan}, {Mellier}, \& {Scoville}}]{cassata14}
{Cassata}, P., {Tasca}, L.~A.~M., {Le Fevre}, O., {et~al.} 2014,
  ArXiv:1403.3693

\bibitem[{{Castellano} {et~al.}(2007){Castellano}, {Salimbeni}, {Trevese},
  {Grazian}, {Pentericci}, {Fiore}, {Fontana}, {Giallongo}, {Santini},
  {Cristiani}, {Nonino}, \& {Vanzella}}]{castellano07}
{Castellano}, M., {Salimbeni}, S., {Trevese}, D., {et~al.} 2007, \apj, 671,
  1497

\bibitem[{{Chiang} {et~al.}(2013){Chiang}, {Overzier}, \&
  {Gebhardt}}]{chiang13_sim}
{Chiang}, Y.-K., {Overzier}, R., \& {Gebhardt}, K. 2013, \apj, 779, 127

\bibitem[{{Chiang} {et~al.}(2014){Chiang}, {Overzier}, \&
  {Gebhardt}}]{chiang14_cosmos}
{Chiang}, Y.-K., {Overzier}, R., \& {Gebhardt}, K. 2014, \apjl, 782, L3

\bibitem[{{Cucciati} {et~al.}(2006){Cucciati}, {Iovino}, {Marinoni}, {Ilbert},
  {Bardelli}, {Franzetti}, {Le F{\`e}vre}, {Pollo}, {Zamorani}, {Cappi},
  {Guzzo}, {McCracken}, {Meneux}, {Scaramella}, {Scodeggio}, {Tresse}, {Zucca},
  {Bottini}, {Garilli}, {Le Brun}, {Maccagni}, {Picat}, {Vettolani},
  {Zanichelli}, {Adami}, {Arnaboldi}, {Arnouts}, {Bolzonella}, {Charlot},
  {Ciliegi}, {Contini}, {Foucaud}, {Gavignaud}, {Marano}, {Mazure}, {Merighi},
  {Paltani}, {Pell{\`o}}, {Pozzetti}, {Radovich}, {Bondi}, {Bongiorno},
  {Busarello}, {de La Torre}, {Gregorini}, {Lamareille}, {Mathez}, {Mellier},
  {Merluzzi}, {Ripepi}, {Rizzo}, {Temporin}, \& {Vergani}}]{cucciati2006}
{Cucciati}, O., {Iovino}, A., {Marinoni}, C., {et~al.} 2006, \aap, 458, 39

\bibitem[{{Cucciati} {et~al.}(2012){Cucciati}, {Tresse}, {Ilbert}, {Le
  F{\`e}vre}, {Garilli}, {Le Brun}, {Cassata}, {Franzetti}, {Maccagni},
  {Scodeggio}, {Zucca}, {Zamorani}, {Bardelli}, {Bolzonella}, {Bielby},
  {McCracken}, {Zanichelli}, \& {Vergani}}]{cucciati12sfrd}
{Cucciati}, O., {Tresse}, L., {Ilbert}, O., {et~al.} 2012, \aap, 539, A31

\bibitem[{{De Lucia} \& {Blaizot}(2007)}]{delucia_blaizot2007}
{De Lucia}, G. \& {Blaizot}, J. 2007, \mnras, 375, 2

\bibitem[{{de Ravel} {et~al.}(2009){de Ravel}, {Le F{\`e}vre}, {Tresse},
  {Bottini}, {Garilli}, {Le Brun}, {Maccagni}, {Scaramella}, {Scodeggio},
  {Vettolani}, {Zanichelli}, {Adami}, {Arnouts}, {Bardelli}, {Bolzonella},
  {Cappi}, {Charlot}, {Ciliegi}, {Contini}, {Foucaud}, {Franzetti},
  {Gavignaud}, {Guzzo}, {Ilbert}, {Iovino}, {Lamareille}, {McCracken},
  {Marano}, {Marinoni}, {Mazure}, {Meneux}, {Merighi}, {Paltani}, {Pell{\`o}},
  {Pollo}, {Pozzetti}, {Radovich}, {Vergani}, {Zamorani}, {Zucca}, {Bondi},
  {Bongiorno}, {Brinchmann}, {Cucciati}, {de La Torre}, {Gregorini}, {Memeo},
  {Perez-Montero}, {Mellier}, {Merluzzi}, \& {Temporin}}]{deravel09}
{de Ravel}, L., {Le F{\`e}vre}, O., {Tresse}, L., {et~al.} 2009, \aap, 498, 379

\bibitem[{{Decarli} {et~al.}(2012){Decarli}, {Walter}, {Yang}, {Carilli},
  {Fan}, {Hennawi}, {Kurk}, {Riechers}, {Rix}, {Strauss}, \&
  {Venemans}}]{decarli12}
{Decarli}, R., {Walter}, F., {Yang}, Y., {et~al.} 2012, \apj, 756, 150

\bibitem[{{Dekel} {et~al.}(2009){Dekel}, {Birnboim}, {Engel}, {Freundlich},
  {Goerdt}, {Mumcuoglu}, {Neistein}, {Pichon}, {Teyssier}, \&
  {Zinger}}]{dekel09}
{Dekel}, A., {Birnboim}, Y., {Engel}, G., {et~al.} 2009, \nat, 457, 451

\bibitem[{{Eke} {et~al.}(1998){Eke}, {Navarro}, \& {Frenk}}]{eke98}
{Eke}, V.~R., {Navarro}, J.~F., \& {Frenk}, C.~S. 1998, \apj, 503, 569

\bibitem[{{Elbaz} {et~al.}(2007){Elbaz}, {Daddi}, {Le Borgne}, {Dickinson},
  {Alexander}, {Chary}, {Starck}, {Brandt}, {Kitzbichler}, {MacDonald},
  {Nonino}, {Popesso}, {Stern}, \& {Vanzella}}]{elbaz2007}
{Elbaz}, D., {Daddi}, E., {Le Borgne}, D., {et~al.} 2007, \aap, 468, 33

\bibitem[{{Elvis} {et~al.}(2009){Elvis}, {Civano}, {Vignali}, {Puccetti},
  {Fiore}, {Cappelluti}, {Aldcroft}, {Fruscione}, {Zamorani}, {Comastri},
  {Brusa}, {Gilli}, {Miyaji}, {Damiani}, {Koekemoer}, {Finoguenov}, {Brunner},
  {Urry}, {Silverman}, {Mainieri}, {Hasinger}, {Griffiths}, {Carollo}, {Hao},
  {Guzzo}, {Blain}, {Calzetti}, {Carilli}, {Capak}, {Ettori}, {Fabbiano},
  {Impey}, {Lilly}, {Mobasher}, {Rich}, {Salvato}, {Sanders}, {Schinnerer},
  {Scoville}, {Shopbell}, {Taylor}, {Taniguchi}, \&
  {Volonteri}}]{elvis09_chandra}
{Elvis}, M., {Civano}, F., {Vignali}, C., {et~al.} 2009, \apjs, 184, 158

\bibitem[{{Ettori} {et~al.}(2009){Ettori}, {Morandi}, {Tozzi}, {Balestra},
  {Borgani}, {Rosati}, {Lovisari}, \& {Terenziani}}]{ettori09}
{Ettori}, S., {Morandi}, A., {Tozzi}, P., {et~al.} 2009, \aap, 501, 61

\bibitem[{{Fassbender} {et~al.}(2011){Fassbender}, {Nastasi}, {B{\"o}hringer},
  {{\v S}uhada}, {Santos}, {Rosati}, {Pierini}, {M{\"u}hlegger}, {Quintana},
  {Schwope}, {Lamer}, {de Hoon}, {Kohnert}, {Pratt}, \& {Mohr}}]{fassbender11}
{Fassbender}, R., {Nastasi}, A., {B{\"o}hringer}, H., {et~al.} 2011, \aap, 527,
  L10

\bibitem[{{Finn} {et~al.}(2005){Finn}, {Zaritsky}, {McCarthy}, {Poggianti},
  {Rudnick}, {Halliday}, {Milvang-Jensen}, {Pell{\'o}}, \& {Simard}}]{finn05}
{Finn}, R.~A., {Zaritsky}, D., {McCarthy}, Jr., D.~W., {et~al.} 2005, \apj,
  630, 206

\bibitem[{{Foley} {et~al.}(2011){Foley}, {Andersson}, {Bazin}, {de Haan},
  {Ruel}, {Ade}, {Aird}, {Armstrong}, {Ashby}, {Bautz}, {Benson}, {Bleem},
  {Bonamente}, {Brodwin}, {Carlstrom}, {Chang}, {Clocchiatti}, {Crawford},
  {Crites}, {Desai}, {Dobbs}, {Dudley}, {Fazio}, {Forman}, {Garmire}, {George},
  {Gladders}, {Gonzalez}, {Halverson}, {High}, {Holder}, {Holzapfel}, {Hoover},
  {Hrubes}, {Jones}, {Joy}, {Keisler}, {Knox}, {Lee}, {Leitch}, {Lueker},
  {Luong-Van}, {Marrone}, {McMahon}, {Mehl}, {Meyer}, {Mohr}, {Montroy},
  {Murray}, {Padin}, {Plagge}, {Pryke}, {Reichardt}, {Rest}, {Ruhl},
  {Saliwanchik}, {Saro}, {Schaffer}, {Shaw}, {Shirokoff}, {Song}, {Spieler},
  {Stalder}, {Stanford}, {Staniszewski}, {Stark}, {Story}, {Stubbs},
  {Vanderlinde}, {Vieira}, {Vikhlinin}, {Williamson}, \&
  {Zenteno}}]{foley11_SZ}
{Foley}, R.~J., {Andersson}, K., {Bazin}, G., {et~al.} 2011, \apj, 731, 86

\bibitem[{{Fumagalli} {et~al.}(2011){Fumagalli}, {Prochaska}, {Kasen}, {Dekel},
  {Ceverino}, \& {Primack}}]{fumagalli11}
{Fumagalli}, M., {Prochaska}, J.~X., {Kasen}, D., {et~al.} 2011, \mnras, 418,
  1796

\bibitem[{{Galametz} {et~al.}(2012){Galametz}, {Stern}, {De Breuck}, {Hatch},
  {Mayo}, {Miley}, {Rettura}, {Seymour}, {Stanford}, \& {Vernet}}]{galametz12}
{Galametz}, A., {Stern}, D., {De Breuck}, C., {et~al.} 2012, \apj, 749, 169

\bibitem[{{Garilli} {et~al.}(2010){Garilli}, {Fumana}, {Franzetti}, {Paioro},
  {Scodeggio}, {Le F{\`e}vre}, {Paltani}, \& {Scaramella}}]{garilli10}
{Garilli}, B., {Fumana}, M., {Franzetti}, P., {et~al.} 2010, \pasp, 122, 827

\bibitem[{{Giavalisco} {et~al.}(2011){Giavalisco}, {Vanzella}, {Salimbeni},
  {Tripp}, {Dickinson}, {Cassata}, {Renzini}, {Guo}, {Ferguson}, {Nonino},
  {Cimatti}, {Kurk}, {Mignoli}, \& {Tang}}]{giavalisco11}
{Giavalisco}, M., {Vanzella}, E., {Salimbeni}, S., {et~al.} 2011, \apj, 743, 95

\bibitem[{{Gobat} {et~al.}(2011){Gobat}, {Daddi}, {Onodera}, {Finoguenov},
  {Renzini}, {Arimoto}, {Bouwens}, {Brusa}, {Chary}, {Cimatti}, {Dickinson},
  {Kong}, \& {Mignoli}}]{gobat11}
{Gobat}, R., {Daddi}, E., {Onodera}, M., {et~al.} 2011, \aap, 526, A133

\bibitem[{{Goerdt} {et~al.}(2010){Goerdt}, {Dekel}, {Sternberg}, {Ceverino},
  {Teyssier}, \& {Primack}}]{goerdt10}
{Goerdt}, T., {Dekel}, A., {Sternberg}, A., {et~al.} 2010, \mnras, 407, 613

\bibitem[{{Goerdt} {et~al.}(2012){Goerdt}, {Dekel}, {Sternberg}, {Gnat}, \&
  {Ceverino}}]{goerdt12}
{Goerdt}, T., {Dekel}, A., {Sternberg}, A., {Gnat}, O., \& {Ceverino}, D. 2012,
  \mnras, 424, 2292

\bibitem[{{Guo} {et~al.}(2011){Guo}, {White}, {Boylan-Kolchin}, {De Lucia},
  {Kauffmann}, {Lemson}, {Li}, {Springel}, \& {Weinmann}}]{guo11_sam}
{Guo}, Q., {White}, S., {Boylan-Kolchin}, M., {et~al.} 2011, \mnras, 413, 101

\bibitem[{{Hatch} {et~al.}(2011){Hatch}, {Kurk}, {Pentericci}, {Venemans},
  {Kuiper}, {Miley}, \& {R{\"o}ttgering}}]{hatch11b}
{Hatch}, N.~A., {Kurk}, J.~D., {Pentericci}, L., {et~al.} 2011, \mnras, 415,
  2993

\bibitem[{{Ilbert} {et~al.}(2006){Ilbert}, {Arnouts}, {McCracken},
  {Bolzonella}, {Bertin}, {Le F{\`e}vre}, {Mellier}, {Zamorani}, {Pell{\`o}},
  {Iovino}, {Tresse}, {Le Brun}, {Bottini}, {Garilli}, {Maccagni}, {Picat},
  {Scaramella}, {Scodeggio}, {Vettolani}, {Zanichelli}, {Adami}, {Bardelli},
  {Cappi}, {Charlot}, {Ciliegi}, {Contini}, {Cucciati}, {Foucaud}, {Franzetti},
  {Gavignaud}, {Guzzo}, {Marano}, {Marinoni}, {Mazure}, {Meneux}, {Merighi},
  {Paltani}, {Pollo}, {Pozzetti}, {Radovich}, {Zucca}, {Bondi}, {Bongiorno},
  {Busarello}, {de La Torre}, {Gregorini}, {Lamareille}, {Mathez}, {Merluzzi},
  {Ripepi}, {Rizzo}, \& {Vergani}}]{ilbert2006_pz}
{Ilbert}, O., {Arnouts}, S., {McCracken}, H.~J., {et~al.} 2006, \aap, 457, 841

\bibitem[{{Ilbert} {et~al.}(2009){Ilbert}, {Capak}, {Salvato}, {Aussel},
  {McCracken}, {Sanders}, {Scoville}, {Kartaltepe}, {Arnouts}, {Le Floc'h},
  {Mobasher}, {Taniguchi}, {Lamareille}, {Leauthaud}, {Sasaki}, {Thompson},
  {Zamojski}, {Zamorani}, {Bardelli}, {Bolzonella}, {Bongiorno}, {Brusa},
  {Caputi}, {Carollo}, {Contini}, {Cook}, {Coppa}, {Cucciati}, {de la Torre},
  {de Ravel}, {Franzetti}, {Garilli}, {Hasinger}, {Iovino}, {Kampczyk},
  {Kneib}, {Knobel}, {Kovac}, {Le Borgne}, {Le Brun}, {F{\`e}vre}, {Lilly},
  {Looper}, {Maier}, {Mainieri}, {Mellier}, {Mignoli}, {Murayama}, {Pell{\`o}},
  {Peng}, {P{\'e}rez-Montero}, {Renzini}, {Ricciardelli}, {Schiminovich},
  {Scodeggio}, {Shioya}, {Silverman}, {Surace}, {Tanaka}, {Tasca}, {Tresse},
  {Vergani}, \& {Zucca}}]{ilbert09}
{Ilbert}, O., {Capak}, P., {Salvato}, M., {et~al.} 2009, \apj, 690, 1236

\bibitem[{{Ilbert} {et~al.}(2013){Ilbert}, {McCracken}, {Le F{\`e}vre},
  {Capak}, {Dunlop}, {Karim}, {Renzini}, {Caputi}, {Boissier}, {Arnouts},
  {Aussel}, {Comparat}, {Guo}, {Hudelot}, {Kartaltepe}, {Kneib}, {Krogager},
  {Le Floc'h}, {Lilly}, {Mellier}, {Milvang-Jensen}, {Moutard}, {Onodera},
  {Richard}, {Salvato}, {Sanders}, {Scoville}, {Silverman}, {Taniguchi},
  {Tasca}, {Thomas}, {Toft}, {Tresse}, {Vergani}, {Wolk}, \&
  {Zirm}}]{ilbert2013}
{Ilbert}, O., {McCracken}, H.~J., {Le F{\`e}vre}, O., {et~al.} 2013, \aap, 556,
  A55

\bibitem[{{Ilbert} {et~al.}(2005){Ilbert}, {Tresse}, {Zucca}, {Bardelli},
  {Arnouts}, {Zamorani}, {Pozzetti}, {Bottini}, {Garilli}, {Le Brun}, {Le
  F{\`e}vre}, {Maccagni}, {Picat}, {Scaramella}, {Scodeggio}, {Vettolani},
  {Zanichelli}, {Adami}, {Arnaboldi}, {Bolzonella}, {Cappi}, {Charlot},
  {Contini}, {Foucaud}, {Franzetti}, {Gavignaud}, {Guzzo}, {Iovino},
  {McCracken}, {Marano}, {Marinoni}, {Mathez}, {Mazure}, {Meneux}, {Merighi},
  {Paltani}, {Pello}, {Pollo}, {Radovich}, {Bondi}, {Bongiorno}, {Busarello},
  {Ciliegi}, {Lamareille}, {Mellier}, {Merluzzi}, {Ripepi}, \&
  {Rizzo}}]{ilbert2005}
{Ilbert}, O., {Tresse}, L., {Zucca}, E., {et~al.} 2005, \aap, 439, 863

\bibitem[{{Kere{\v s}} {et~al.}(2005){Kere{\v s}}, {Katz}, {Weinberg}, \&
  {Dav{\'e}}}]{keres05}
{Kere{\v s}}, D., {Katz}, N., {Weinberg}, D.~H., \& {Dav{\'e}}, R. 2005,
  \mnras, 363, 2

\bibitem[{{Kimm} {et~al.}(2011){Kimm}, {Slyz}, {Devriendt}, \&
  {Pichon}}]{kimm11}
{Kimm}, T., {Slyz}, A., {Devriendt}, J., \& {Pichon}, C. 2011, \mnras, 413, L51

\bibitem[{{Kodama} {et~al.}(2007){Kodama}, {Tanaka}, {Kajisawa}, {Kurk},
  {Venemans}, {De Breuck}, {Vernet}, \& {Lidman}}]{kodama07}
{Kodama}, T., {Tanaka}, I., {Kajisawa}, M., {et~al.} 2007, \mnras, 377, 1717

\bibitem[{{Koekemoer} {et~al.}(2007){Koekemoer}, {Aussel}, {Calzetti}, {Capak},
  {Giavalisco}, {Kneib}, {Leauthaud}, {Le F{\`e}vre}, {McCracken}, {Massey},
  {Mobasher}, {Rhodes}, {Scoville}, \& {Shopbell}}]{koekemoer07_HST}
{Koekemoer}, A.~M., {Aussel}, H., {Calzetti}, D., {et~al.} 2007, \apjs, 172,
  196

\bibitem[{{Kurk} {et~al.}(2009){Kurk}, {Cimatti}, {Zamorani}, {Halliday},
  {Mignoli}, {Pozzetti}, {Daddi}, {Rosati}, {Dickinson}, {Bolzonella},
  {Cassata}, {Renzini}, {Franceschini}, {Rodighiero}, \& {Berta}}]{kurk09}
{Kurk}, J., {Cimatti}, A., {Zamorani}, G., {et~al.} 2009, \aap, 504, 331

\bibitem[{{Le F{\`e}vre} {et~al.}(2013){Le F{\`e}vre}, {Cassata}, {Cucciati},
  {Garilli}, {Ilbert}, {Le Brun}, {Maccagni}, {Moreau}, {Scodeggio}, {Tresse},
  {Zamorani}, {Adami}, {Arnouts}, {Bardelli}, {Bolzonella}, {Bondi},
  {Bongiorno}, {Bottini}, {Cappi}, {Charlot}, {Ciliegi}, {Contini}, {de la
  Torre}, {Foucaud}, {Franzetti}, {Gavignaud}, {Guzzo}, {Iovino}, {Lemaux},
  {L{\'o}pez-Sanjuan}, {McCracken}, {Marano}, {Marinoni}, {Mazure}, {Mellier},
  {Merighi}, {Merluzzi}, {Paltani}, {Pell{\`o}}, {Pollo}, {Pozzetti},
  {Scaramella}, {Tasca}, {Vergani}, {Vettolani}, {Zanichelli}, \&
  {Zucca}}]{lefevre2013a}
{Le F{\`e}vre}, O., {Cassata}, P., {Cucciati}, O., {et~al.} 2013, \aap, 559,
  A14

\bibitem[{{Le F{\`e}vre} {et~al.}(2003){Le F{\`e}vre}, {Saisse}, {Mancini},
  {Brau-Nogue}, {Caputi}, {Castinel}, {D'Odorico}, {Garilli}, {Kissler-Patig},
  {Lucuix}, {Mancini}, {Pauget}, {Sciarretta}, {Scodeggio}, {Tresse}, \&
  {Vettolani}}]{lefevre2003}
{Le F{\`e}vre}, O., {Saisse}, M., {Mancini}, D., {et~al.} 2003, in Instrument
  Design and Performance for Optical/Infrared Ground-based Telescopes. Edited
  by Iye, Masanori; Moorwood, Alan F. M. Proceedings of the SPIE, Volume 4841,
  pp. 1670-1681 (2003)., 1670--1681

\bibitem[{{Le Fevre} {et~al.}(2014){Le Fevre}, {Tasca}, {Cassata}, {Garilli},
  {Le Brun}, {Maccagni}, {Pentericci}, {Thomas}, {Vanzella}, {Zamorani},
  {Zucca}, {Amorin}, {Bardelli}, {Capak}, {Cassara}, {Castellano}, {Cimatti},
  {Cuby}, {Cucciati}, {de la Torre}, {Durkalec}, {Fontana}, {Giavalisco},
  {Grazian}, {Hathi}, {Ilbert}, {Lemaux}, {Moreau}, {Paltani}, {Ribeiro},
  {Salvato}, {Schaerer}, {Scodeggio}, {Sommariva}, {Talia}, {Taniguchi},
  {Tresse}, {Vergani}, {Wang}, {Charlot}, {Contini}, {Fotopoulo},
  {Lopez-Sanjuan}, {Mellier}, \& {Scoville}}]{lefevre14}
{Le Fevre}, O., {Tasca}, L.~A.~M., {Cassata}, P., {et~al.} 2014,
  ArXiv:1403.3938

\bibitem[{{Lemaux} {et~al.}(2014){Lemaux}, {Cucciati}, {Tasca}, {Le F{\`e}vre},
  {Zamorani}, {Cassata}, {Garilli}, {Le Brun}, {Maccagni}, {Pentericci},
  {Thomas}, {Vanzella}, {Zucca}, {Amorin}, {Bardelli}, {Capak}, {Cassar{\`a}},
  {Castellano}, {Cimatti}, {Cuby}, {de la Torre}, {Durkalec}, {Fontana},
  {Giavalisco}, {Grazian}, {Hathi}, {Ilbert}, {Moreau}, {Paltani}, {Ribeiro},
  {Salvato}, {Schaerer}, {Scodeggio}, {Sommariva}, {Talia}, {Taniguchi},
  {Tresse}, {Vergani}, {Wang}, {Charlot}, {Contini}, {Fotopoulou}, {Gal},
  {Kocevski}, {L{\'o}pez-Sanjuan}, {Lubin}, {Mellier}, {Sadibekova}, \&
  {Scoville}}]{lemaux14}
{Lemaux}, B.~C., {Cucciati}, O., {Tasca}, L.~A.~M., {et~al.} 2014,
  ArXiv:1403.4230

\bibitem[{{Lemaux} {et~al.}(2009){Lemaux}, {Lubin}, {Sawicki}, {Martin},
  {Lagattuta}, {Gal}, {Kocevski}, {Fassnacht}, \& {Squires}}]{lemaux09}
{Lemaux}, B.~C., {Lubin}, L.~M., {Sawicki}, M., {et~al.} 2009, \apj, 700, 20

\bibitem[{{L{\'o}pez-Sanjuan} {et~al.}(2011){L{\'o}pez-Sanjuan}, {Le
  F{\`e}vre}, {de Ravel}, {Cucciati}, {Ilbert}, {Tresse}, {Bardelli},
  {Bolzonella}, {Contini}, {Garilli}, {Guzzo}, {Maccagni}, {McCracken},
  {Mellier}, {Pollo}, {Vergani}, \& {Zucca}}]{lopez_sanjuan11_vvds}
{L{\'o}pez-Sanjuan}, C., {Le F{\`e}vre}, O., {de Ravel}, L., {et~al.} 2011,
  \aap, 530, A20

\bibitem[{{L{\'o}pez-Sanjuan} {et~al.}(2013){L{\'o}pez-Sanjuan}, {Le
  F{\`e}vre}, {Tasca}, {Epinat}, {Amram}, {Contini}, {Garilli},
  {Kissler-Patig}, {Moultaka}, {Paioro}, {Perret}, {Queyrel}, {Tresse},
  {Vergani}, \& {Divoy}}]{lopez_sanjuan13_massiv}
{L{\'o}pez-Sanjuan}, C., {Le F{\`e}vre}, O., {Tasca}, L.~A.~M., {et~al.} 2013,
  \aap, 553, A78

\bibitem[{{Lutz} {et~al.}(2011){Lutz}, {Poglitsch}, {Altieri}, {Andreani},
  {Aussel}, {Berta}, {Bongiovanni}, {Brisbin}, {Cava}, {Cepa}, {Cimatti},
  {Daddi}, {Dominguez-Sanchez}, {Elbaz}, {F{\"o}rster Schreiber}, {Genzel},
  {Grazian}, {Gruppioni}, {Harwit}, {Le Floc'h}, {Magdis}, {Magnelli},
  {Maiolino}, {Nordon}, {P{\'e}rez Garc{\'{\i}}a}, {Popesso}, {Pozzi},
  {Riguccini}, {Rodighiero}, {Saintonge}, {Sanchez Portal}, {Santini}, {Shao},
  {Sturm}, {Tacconi}, {Valtchanov}, {Wetzstein}, \& {Wieprecht}}]{lutz11_PEP}
{Lutz}, D., {Poglitsch}, A., {Altieri}, B., {et~al.} 2011, \aap, 532, A90

\bibitem[{{Madau} {et~al.}(1996){Madau}, {Ferguson}, {Dickinson}, {Giavalisco},
  {Steidel}, \& {Fruchter}}]{madau1996}
{Madau}, P., {Ferguson}, H.~C., {Dickinson}, M.~E., {et~al.} 1996, \mnras, 283,
  1388

\bibitem[{{Matsuda} {et~al.}(2009){Matsuda}, {Nakamura}, {Morimoto}, {Smail},
  {De Breuck}, {Ohta}, {Kodama}, {Inoue}, {Hayashino}, {Kousai}, {Nakamura},
  {Horie}, {Yamada}, {Kitamura}, {Saito}, {Taniguchi}, {Tanaka}, \&
  {Hibon}}]{matsuda09}
{Matsuda}, Y., {Nakamura}, Y., {Morimoto}, N., {et~al.} 2009, \mnras, 400, L66

\bibitem[{{McCracken} {et~al.}(2012){McCracken}, {Milvang-Jensen}, {Dunlop},
  {Franx}, {Fynbo}, {Le F{\`e}vre}, {Holt}, {Caputi}, {Goranova}, {Buitrago},
  {Emerson}, {Freudling}, {Hudelot}, {L{\'o}pez-Sanjuan}, {Magnard}, {Mellier},
  {M{\o}ller}, {Nilsson}, {Sutherland}, {Tasca}, \& {Zabl}}]{mccracken12}
{McCracken}, H.~J., {Milvang-Jensen}, B., {Dunlop}, J., {et~al.} 2012, \aap,
  544, A156

\bibitem[{{Miley} {et~al.}(2004){Miley}, {Overzier}, {Tsvetanov}, {Bouwens},
  {Ben{\'{\i}}tez}, {Blakeslee}, {Ford}, {Illingworth}, {Postman}, {Rosati},
  {Clampin}, {Hartig}, {Zirm}, {R{\"o}ttgering}, {Venemans}, {Ardila},
  {Bartko}, {Broadhurst}, {Brown}, {Burrows}, {Cheng}, {Cross}, {De Breuck},
  {Feldman}, {Franx}, {Golimowski}, {Gronwall}, {Infante}, {Martel},
  {Menanteau}, {Meurer}, {Sirianni}, {Kimble}, {Krist}, {Sparks}, {Tran},
  {White}, \& {Zheng}}]{miley04}
{Miley}, G.~K., {Overzier}, R.~A., {Tsvetanov}, Z.~I., {et~al.} 2004, \nat,
  427, 47

\bibitem[{{Ouchi} {et~al.}(2005){Ouchi}, {Shimasaku}, {Akiyama}, {Sekiguchi},
  {Furusawa}, {Okamura}, {Kashikawa}, {Iye}, {Kodama}, {Saito}, {Sasaki},
  {Simpson}, {Takata}, {Yamada}, {Yamanoi}, {Yoshida}, \& {Yoshida}}]{Ouchi05}
{Ouchi}, M., {Shimasaku}, K., {Akiyama}, M., {et~al.} 2005, \apjl, 620, L1

\bibitem[{{Ouchi} {et~al.}(2003){Ouchi}, {Shimasaku}, {Furusawa}, {Miyazaki},
  {Doi}, {Hamabe}, {Hayashino}, {Kimura}, {Kodaira}, {Komiyama}, {Matsuda},
  {Miyazaki}, {Nakata}, {Okamura}, {Sekiguchi}, {Shioya}, {Tamura},
  {Taniguchi}, {Yagi}, \& {Yasuda}}]{Ouchi03}
{Ouchi}, M., {Shimasaku}, K., {Furusawa}, H., {et~al.} 2003, \apj, 582, 60

\bibitem[{{Pentericci} {et~al.}(2000){Pentericci}, {Kurk}, {R{\"o}ttgering},
  {Miley}, {van Breugel}, {Carilli}, {Ford}, {Heckman}, {McCarthy}, \&
  {Moorwood}}]{pentericci00}
{Pentericci}, L., {Kurk}, J.~D., {R{\"o}ttgering}, H.~J.~A., {et~al.} 2000,
  \aap, 361, L25

\bibitem[{{Salimbeni} {et~al.}(2009){Salimbeni}, {Castellano}, {Pentericci},
  {Trevese}, {Fiore}, {Grazian}, {Fontana}, {Giallongo}, {Boutsia},
  {Cristiani}, {de Santis}, {Gallozzi}, {Menci}, {Nonino}, {Paris}, {Santini},
  \& {Vanzella}}]{salimbeni09}
{Salimbeni}, S., {Castellano}, M., {Pentericci}, L., {et~al.} 2009, \aap, 501,
  865

\bibitem[{{Scodeggio} {et~al.}(2005){Scodeggio}, {Franzetti}, {Garilli},
  {Zanichelli}, {Paltani}, {Maccagni}, {Bottini}, {Le Brun}, {Contini},
  {Scaramella}, {Adami}, {Bardelli}, {Zucca}, {Tresse}, {Ilbert}, {Foucaud},
  {Iovino}, {Merighi}, {Zamorani}, {Gavignaud}, {Rizzo}, {McCracken}, {Le
  F{\`e}vre}, {Picat}, {Vettolani}, {Arnaboldi}, {Arnouts}, {Bolzonella},
  {Cappi}, {Charlot}, {Ciliegi}, {Guzzo}, {Marano}, {Marinoni}, {Mathez},
  {Mazure}, {Meneux}, {Pell{\`o}}, {Pollo}, {Pozzetti}, \&
  {Radovich}}]{scodeggio05}
{Scodeggio}, M., {Franzetti}, P., {Garilli}, B., {et~al.} 2005, \pasp, 117,
  1284

\bibitem[{{Scoville} {et~al.}(2013){Scoville}, {Arnouts}, {Aussel}, {Benson},
  {Bongiorno}, {Bundy}, {Calvo}, {Capak}, {Carollo}, {Civano}, {Dunlop},
  {Elvis}, {Faisst}, {Finoguenov}, {Fu}, {Giavalisco}, {Guo}, {Ilbert},
  {Iovino}, {Kajisawa}, {Kartaltepe}, {Leauthaud}, {Le F{\`e}vre}, {LeFloch},
  {Lilly}, {Liu}, {Manohar}, {Massey}, {Masters}, {McCracken}, {Mobasher},
  {Peng}, {Renzini}, {Rhodes}, {Salvato}, {Sanders}, {Sarvestani}, {Scarlata},
  {Schinnerer}, {Sheth}, {Shopbell}, {Smol{\v c}i{\'c}}, {Taniguchi}, {Taylor},
  {White}, \& {Yan}}]{scoville13}
{Scoville}, N., {Arnouts}, S., {Aussel}, H., {et~al.} 2013, \apjs, 206, 3

\bibitem[{{Scoville} {et~al.}(2007){Scoville}, {Aussel}, {Brusa}, {Capak},
  {Carollo}, {Elvis}, {Giavalisco}, {Guzzo}, {Hasinger}, {Impey}, {Kneib},
  {LeFevre}, {Lilly}, {Mobasher}, {Renzini}, {Rich}, {Sanders}, {Schinnerer},
  {Schminovich}, {Shopbell}, {Taniguchi}, \& {Tyson}}]{scoville2007_COSMOS}
{Scoville}, N., {Aussel}, H., {Brusa}, M., {et~al.} 2007, \apjs, 172, 1

\bibitem[{{Spitler} {et~al.}(2012){Spitler}, {Labb{\'e}}, {Glazebrook},
  {Persson}, {Monson}, {Papovich}, {Tran}, {Poole}, {Quadri}, {van Dokkum},
  {Kelson}, {Kacprzak}, {McCarthy}, {Murphy}, {Straatman}, \&
  {Tilvi}}]{spitler12}
{Spitler}, L.~R., {Labb{\'e}}, I., {Glazebrook}, K., {et~al.} 2012, \apjl, 748,
  L21

\bibitem[{{Springel} {et~al.}(2005){Springel}, {White}, {Jenkins}, {Frenk},
  {Yoshida}, {Gao}, {Navarro}, {Thacker}, {Croton}, {Helly}, {Peacock}, {Cole},
  {Thomas}, {Couchman}, {Evrard}, {Colberg}, \& {Pearce}}]{springel2005_MILL}
{Springel}, V., {White}, S.~D.~M., {Jenkins}, A., {et~al.} 2005, \nat, 435, 629

\bibitem[{{Stanford} {et~al.}(2012){Stanford}, {Brodwin}, {Gonzalez},
  {Zeimann}, {Stern}, {Dey}, {Eisenhardt}, {Snyder}, \& {Mancone}}]{stanford12}
{Stanford}, S.~A., {Brodwin}, M., {Gonzalez}, A.~H., {et~al.} 2012, \apj, 753,
  164

\bibitem[{{Steidel} {et~al.}(1998){Steidel}, {Adelberger}, {Dickinson},
  {Giavalisco}, {Pettini}, \& {Kellogg}}]{steidel98}
{Steidel}, C.~C., {Adelberger}, K.~L., {Dickinson}, M., {et~al.} 1998, \apj,
  492, 428

\bibitem[{{Steidel} {et~al.}(2005){Steidel}, {Adelberger}, {Shapley}, {Erb},
  {Reddy}, \& {Pettini}}]{steidel05}
{Steidel}, C.~C., {Adelberger}, K.~L., {Shapley}, A.~E., {et~al.} 2005, \apj,
  626, 44

\bibitem[{{Steidel} {et~al.}(2000){Steidel}, {Adelberger}, {Shapley},
  {Pettini}, {Dickinson}, \& {Giavalisco}}]{steidel00}
{Steidel}, C.~C., {Adelberger}, K.~L., {Shapley}, A.~E., {et~al.} 2000, \apj,
  532, 170

\bibitem[{{Sunyaev} \& {Zeldovich}(1980)}]{sunyaev_zeldovich1980}
{Sunyaev}, R.~A. \& {Zeldovich}, I.~B. 1980, \araa, 18, 537

\bibitem[{{Sunyaev} \& {Zeldovich}(1972)}]{sunyaev_zeldovich1972}
{Sunyaev}, R.~A. \& {Zeldovich}, Y.~B. 1972, Comments on Astrophysics and Space
  Physics, 4, 173

\bibitem[{{Taniguchi} {et~al.}(2007){Taniguchi}, {Scoville}, {Murayama},
  {Sanders}, {Mobasher}, {Aussel}, {Capak}, {Ajiki}, {Miyazaki}, {Komiyama},
  {Shioya}, {Nagao}, {Sasaki}, {Koda}, {Carilli}, {Giavalisco}, {Guzzo},
  {Hasinger}, {Impey}, {LeFevre}, {Lilly}, {Renzini}, {Rich}, {Schinnerer},
  {Shopbell}, {Kaifu}, {Karoji}, {Arimoto}, {Okamura}, \& {Ohta}}]{taniguchi07}
{Taniguchi}, Y., {Scoville}, N., {Murayama}, T., {et~al.} 2007, \apjs, 172, 9

\bibitem[{{Tasca} {et~al.}(2014){Tasca}, {Le F{\`e}vre}, {L{\'o}pez-Sanjuan},
  {Wang}, {Cassata}, {Garilli}, {Ilbert}, {Le Brun}, {Lemaux}, {Maccagni},
  {Tresse}, {Bardelli}, {Contini}, {Charlot}, {Cucciati}, {Fontana},
  {Giavalisco}, {Kneib}, {Salvato}, {Taniguchi}, {Vergani}, {Zamorani}, \&
  {Zucca}}]{tasca14_merger}
{Tasca}, L.~A.~M., {Le F{\`e}vre}, O., {L{\'o}pez-Sanjuan}, C., {et~al.} 2014,
  \aap, 565, A10

\bibitem[{{Toshikawa} {et~al.}(2012){Toshikawa}, {Kashikawa}, {Ota},
  {Morokuma}, {Shibuya}, {Hayashi}, {Nagao}, {Jiang}, {Malkan}, {Egami},
  {Shimasaku}, {Motohara}, \& {Ishizaki}}]{toshikawa12}
{Toshikawa}, J., {Kashikawa}, N., {Ota}, K., {et~al.} 2012, \apj, 750, 137

\bibitem[{{Tresse} {et~al.}(1999){Tresse}, {Maddox}, {Loveday}, \&
  {Singleton}}]{tresse99}
{Tresse}, L., {Maddox}, S., {Loveday}, J., \& {Singleton}, C. 1999, \mnras,
  310, 262

\bibitem[{{Venemans} {et~al.}(2007){Venemans}, {R{\"o}ttgering}, {Miley}, {van
  Breugel}, {de Breuck}, {Kurk}, {Pentericci}, {Stanford}, {Overzier}, {Croft},
  \& {Ford}}]{venemans07}
{Venemans}, B.~P., {R{\"o}ttgering}, H.~J.~A., {Miley}, G.~K., {et~al.} 2007,
  \aap, 461, 823

\bibitem[{{Vikhlinin} {et~al.}(2009){Vikhlinin}, {Kravtsov}, {Burenin},
  {Ebeling}, {Forman}, {Hornstrup}, {Jones}, {Murray}, {Nagai}, {Quintana}, \&
  {Voevodkin}}]{vikhlinin09}
{Vikhlinin}, A., {Kravtsov}, A.~V., {Burenin}, R.~A., {et~al.} 2009, \apj, 692,
  1060

\bibitem[{{Voronoi}(1908)}]{voronoi1908}
{Voronoi}, G.~F. 1908, J. Reine Angew. Math., 134, 198

\bibitem[{{Williams} {et~al.}(2009){Williams}, {Quadri}, {Franx}, {van Dokkum},
  \& {Labb{\'e}}}]{williams09_UVJ}
{Williams}, R.~J., {Quadri}, R.~F., {Franx}, M., {van Dokkum}, P., \&
  {Labb{\'e}}, I. 2009, \apj, 691, 1879

\end{thebibliography}

\appendix


\section{Testing the robustness of the absorption line}\label{test_abs}

We implemented an automatic procedure to measure 
the EW of all the absorption features (real or not) in a homogeneous way 
in the 1000
stacked spectra mentioned at the end of Sect.~\ref{background}.
Specifically, we performed the following analysis at each pixel (at
position $\lambda_i$) of the stacked spectrum. We define the
continuum $f_{c,i}$ around $\lambda_i$ as the mean value of the flux
of the coadded spectrum in the two intervals $\lambda_i-280 < \lambda
< \lambda_i-80 $\AA~ and $\lambda_i+80 < \lambda < \lambda_i+280
$\AA. To compute the mean, we use only the pixels where $N_{s,i}\geq
4$ (see Sect.~\ref{background} for the definition of $N_{s,i}$). If
the flux $f_i$ at $\lambda_i$ is above $f_{c,i}$, there is no 
absorption line at that position, so we move to the next
$\lambda_i$. If $f_i$ is below $f_{c,i}$, we compute the $\EW_i$ of
the possible line centred at $\lambda_i$, considering as `line' the
spectrum in the range $\lambda_i-30 < \lambda < \lambda_i+30$ \AA.

The output of this procedure is a value of EW$_i$ for each $\lambda_i$
with $f_i < f_{c,i}$ in the coadded spectrum. By definition, $\EW_i$
is positive for absorption features and negative for emission
features. In our analysis, we keep only the $\EW {\rm s}_i$ that
satisfy the following requirements:

\begin{itemize}

\item[-] $\EW_i$ must be positive.

\item[-] The continuum around the line must be entirely included in the 
observed spectrum, so we need $(\lambda_i-280)>3600$\AA~ and  
$(\lambda_i+280)<9350$\AA.

\item[-] The flux at $\lambda_i$ must be a local minimum,
i.e. $f_i<f_{i-1}$ and $f_i<f_{i+1}$.  In the majority of cases, $f_i$
is less than $f_{c,i}$ only because of noise fluctuations around the
continuum, and most probably $f_{i-1}$ and $f_{i+1}$ will be above the
continuum. We want to keep all these cases. In the case of absorption
features larger than one pixel and centred at a given $\lambda_c$,
$f_i$ will be below the continuum for $\lambda_i$ included in a given
range $[\lambda_c-L, \lambda_c+L]$, where $2L$ is the width of the
absorption feature at the level of the continuum. Our automatic
procedure computes the EW centred at all
possible $\lambda_i$ included in this range, and the output is a series of
positive $\EW_i$ with (most probably) a (local) maximum EW at
$\lambda_c$. We retain only $\EW_i$ values where $f_i$ is a local minimum, 
thus normally we keep the largest possible EW for each absorption feature.

\item[-] $N_{s,i}$ in the entire range 
$\lambda_i-30 < \lambda < \lambda_i+30$ \AA~ must be $\geq 5$.

\item[-] The continuum flux $f_{c,i}$ must be computed in a range of
at least 100\AA. As described above, $f_{c,i}$ is measured using two
intervals of $\sim 200$\AA~ each, but using only the fluxes $f_i$
where $N_{s,i}\geq 4$. In some case, this threshold shrinks the
wavelength range on which $f_{c,i}$ is computed, especially at
$\lambda > 6000$, where the sky level in VUDS spectra is on average
quite high because of fringing.

\item[-] The continuum measured on the left ($f_{c,i}^l$) and on the
right ($f_{c,i}^r$) of $\lambda_i$ must be similar. Namely,
max($f_{c,i}^l / f_{c,i}^r$, $f_{c,i}^r / f_{c,i}^l$) $\leq 1.5$. In
this way we avoid absorption features that are too close to a break in the
continuum, a situation difficult to be treated properly when measuring
EW automatically for all features, without knowing their
position and the shape of the continuum around them in advance.

\end{itemize}

\noindent The EW distribution used for the analysis at the end of
Sec.~\ref{background} is computed using all the EWs in the 1000
coadded spectra that satisfy the conditions listed above. We
verified that these conditions are also satisfied by the absorption
line at $\lambda= 4736$\AA~ in the stacked spectra shown
in Fig.~\ref{stacked_spec}.

\end{document}